\documentclass[conference,10pt]{IEEEtran}
\IEEEoverridecommandlockouts
\usepackage{cite}
\usepackage{amsmath,amssymb,amsfonts}
\usepackage{algorithmic}
\usepackage{graphicx}
\usepackage{textcomp}
\usepackage{xcolor}
\def\BibTeX{{\rm B\kern-.05em{\sc i\kern-.025em b}\kern-.08em
    T\kern-.1667em\lower.7ex\hbox{E}\kern-.125emX}}
    
\AtBeginDocument{%
  \providecommand\BibTeX{{%
    \normalfont B\kern-0.5em{\scshape i\kern-0.25em b}\kern-0.8em\TeX}}}

\usepackage{subcaption}
\usepackage{tikz}
\usetikzlibrary{matrix}
\usetikzlibrary{shapes.geometric}
\usetikzlibrary{arrows.meta,shapes,snakes}
\usetikzlibrary{positioning}
\usepackage{bm}
\usepackage{cleveref}
\usepackage{enumitem}
\usepackage{relsize}
\usepackage{multirow}
\usepackage{nicefrac}
\usepackage{amsthm}
\usepackage{booktabs}
\usepackage{comment}
\usepackage{floatrow}

\newcommand{\abs}[1]{\left| #1 \right|}
\usepackage{textcomp}
\newcommand{\textapprox}{\!\raisebox{0.5ex}{\texttildelow}}

\newcommand{\change}[1]{#1}
\theoremstyle{definition}
\newtheorem{definition}{Definition}
\newtheorem{example}{Example}

\begin{document}

\title{SpectralFly: Ramanujan Graphs as Flexible and Efficient Interconnection Networks}

\author{
  \IEEEauthorblockN{
  Stephen Young\IEEEauthorrefmark{1}\textsuperscript{\textsection},
  Sinan Aksoy\IEEEauthorrefmark{1}\textsuperscript{\textsection},
  Jesun Firoz\IEEEauthorrefmark{1}\textsuperscript{\textsection},
  Roberto Gioiosa\IEEEauthorrefmark{1},
  Tobias Hagge\IEEEauthorrefmark{1},
  Mark Kempton\IEEEauthorrefmark{2},\\
  Juan Escobedo\IEEEauthorrefmark{1},
  Mark Raugas\IEEEauthorrefmark{1}
  }
  \IEEEauthorblockA{
    \IEEEauthorrefmark{1}Pacific Northwest National Laboratory,
    \IEEEauthorrefmark{2}Brigham Young University
  }
  \IEEEauthorblockA{
    \IEEEauthorrefmark{1}\{\{first name\}.\{last name\}\}@pnnl.gov,
    \IEEEauthorrefmark{2}mkempton@mathematics.byu.edu
  }
}

\thispagestyle{plain}
\pagestyle{plain}

\maketitle
\begingroup\renewcommand\thefootnote{\textsection}
\footnotetext{Young, Aksoy, and Firoz contributed equally to this work.}
\endgroup

\begin{abstract}
In recent years, graph theoretic considerations have become increasingly important in the design of HPC interconnection topologies. One approach is to seek optimal or near-optimal families of graphs with respect to a particular graph theoretic property, such as diameter. In this work, we consider topologies which optimize the spectral gap. We study a novel HPC topology, SpectralFly, designed around the Ramanujan graph construction of Lubotzky, Phillips, and Sarnak (LPS). We show combinatorial properties, such as diameter, bisection bandwidth, average path length, and resilience to link failure, of SpectralFly topologies are better than, or comparable to, similarly constrained DragonFly, SlimFly, and BundleFly topologies. Additionally, we simulate the performance of SpectralFly on a representative sample of micro-benchmarks using the Structure Simulation Toolkit Macroscale Element Library simulator and study cost-minimizing layouts, demonstrating considerable benefit of the SpectralFly topology.
\end{abstract}


\section{Introduction}


In recent years, a deluge of different interconnection networks have been proposed to address the critical role of communication in modern HPC systems \cite{besta2014slim,lei2020bundlefly,hawkins2007data,kim2008technology,shpiner2017dragonfly+,singla2012jellyfish,valadarsky2015xpander}. 
To efficiently and robustly enable communication, many of these topologies are designed to exhibit a plethora of beneficial structural properties. An ideal network will have low endpoint-to-endpoint latency, resiliency to link failures, high bisection bandwidth to avoid congestion -- all while maintaining low-system cost. 
Researchers have employed the language of graph theory to formalize and quantify such properties. In this way, a number of graph statistics --  such as diameter, average distance, vertex and edge-connectivity, and dense bipartitions -- are well-known to be critically important to the performance of the computing system. However, constructing a graph topology simultaneously optimizing these criteria is challenging.

One approach is to focus on finding families of graphs that are extremal with respect to a particular property, with the hope that optimization of that property guarantees acceptably good, if not optimal, behavior with respect to the others. 
For example, the SlimFly topology \cite{besta2014slim} was proposed with the claim ``it's \emph{ALL} about the diameter." Specifically, the authors argued that graphs which minimize the diameter while simultaneously maximizing the number of vertices for a given radix also exhibit other good properties, such as resilience to link failure and high bisection bandwidth. However, how to construct such topologies (and whether they even exist) remains a challenging and ongoing topic in mathematics~\cite{miller2012moore}. Indeed, the SlimFly topology is only made possible by a sophisticated algebraic construction by McKay, Miller and \v{S}ir\'{a}\v{n} \cite{mckay1998note}, which has nearly-optimal size with respect to degree-diameter condition. SlimFly is far from the only proposed topology to take an extremal approach; the well-known DragonFly \cite{kim2008technology} and associated variants aim to maximize performance while minimizing system cost, utilizing a group of high-radix routers as a virtual router to increase the network's effective radix. And more recently, a related topology called BundleFly~\cite{lei2020bundlefly} expands and adapts the SlimFly for use with multicore fiber systems. 

In this work, we propose that utilizing a graph construction which optimizes the {\it spectral gap} -- the difference between the largest two eigenvalues of the adjacency matrix -- yields a broad family of flexible, balanced, low-cost, and congestion-avoiding topologies. We call this family of topologies SpectralFly, as they are examples of Ramanujan graphs which have optimal spectral gap. As we explain, the spectral gap is a highly nuanced and far-reaching indicator of graph structure, controlling diameter, average distance, fault tolerance, neighborhood expansion, and bisection bandwidth, among others. In comparison to SlimFly, we show SpectralFly makes marginal sacrifices in terms of diameter and average shortest path length, while offering comparable or sometimes significantly better properties, particularly in the case of bisection bandwidth and related properties involving bottlenecks. 
While no single topology can be optimal in every regard, our results show SpectralFly is extremely competitive for many key structural factors, making it a good fit for a variety of workloads. 

Our work is organized as follows: in Section \ref{sec:eigs}, we provide a brief overview of graph eigenvalues, the spectral gap, and the Ramanujan property, establishing the importance of the spectral gap as a consideration in HPC interconnection network design. In Section \ref{sec:const}, we introduce the particular family of Ramanujan graphs we utilize, known as LPS graphs, providing the necessary definitions and examples, and highlighting some key LPS graph properties. Then, in Section \ref{sec:structProp}, we study structural properties of SpectralFly in comparison with SlimFly, BundleFly and DragonFly, across 5 classes of topology sizes which range from roughly 100 vertices to almost 10K vertices. Additionally, we also 
examine the resilience of these properties under varying levels of edge failures. Finally, in Sections \ref{sec:route}-\ref{sec:sim}, we validate the utility of the structural advantages of SpectralFly by performing 
simulations and experiments using the Structural Simulation Toolkit Macroscale Element Library (SST/macro). We define our 
routing algorithms in Section \ref{sec:route}, 
and evaluate several 
micro-benchmarks with different topologies in Section \ref{sec:sim}. 

Throughout, we use standard graph theory terminology and consider only undirected graphs $G=(V,E)$, where $V$ is a set of elements called {\it vertices} and $E$ is a set of unordered pairs of elements of $V$ called {\it edges}. \change{In the context of interconnection networks, vertices represent routers, and edges represent bidirectional links.} The degree of a vertex is the number of edges to which it belongs; we call a graph $k$-regular if each vertex has degree $k$ and sometimes refer to $k$ as the radix of the graph. 

\setlength{\abovedisplayskip}{2.0pt}
\setlength{\belowdisplayskip}{2.0pt}

\section{Eigenvalues, expansion, and the Ramanujan property}\label{sec:eigs}

Graph eigenvalues capture a plethora of network properties critical to the design and function of interconnection networks. Diameter, bisection bandwidth, fault tolerance, average shortest path length, and other structural properties are controlled by eigenvalues; see \cite{aksoy2020ramanujan,Mohar:Eigenvalues} for a survey. \change{Here, we highlight graph theoretic results establishing these connections in order to explain why Ramanujan graphs possess superior structural properties for interconnection network design. }

Many such properties are controlled by a {\it single} eigenvalue: if $G$ is a $k$-regular graph with adjacency matrix $A$, this eigenvalue, denoted $\lambda(G)$, is the largest magnitude eigenvalue of $A$ not equal to $\pm k$. The difference between the two largest adjacency eigenvalues is sometimes called the ``spectral gap". In case of $k$-regular graphs, this is the difference between the second largest eigenvalue and $k$, as $k$ is always the largest eigenvalue. As we will soon explain, Ramanujan graphs ``optimize" the $\lambda(G)$ and hence have large spectral gap. 

Perhaps the most important property for our purposes here, $\lambda(G)$ controls the expansion properties of the graph. Loosely speaking, expansion means every ``not too large" set of vertices has a ``not too small" set of neighbors. The vertex isoperimetric number of a graph, $h(G)$, is one way of formalizing this:
\[
h(G)=\min_{\substack{X \subseteq V(G) \\ 2|X|\leq |V(G)|}}\frac{|\partial X|}{|X|},
\]
where $\partial X$ denotes the neighbors of $X$ that are not in $X$. \change{Thus, larger values of $h(G)$ suggest better expansion properties.}

The vertex isopermetric number, as well as related variants, are closely linked to $\lambda(G)$. In particular, Tanner \cite{Tanner1984} proved a lower bound on $h(G)$ for $k$-regular graphs in terms of this eigenvalue, while Alon and Milman \cite{Alon1985} gave an upper bound. Such results suggest it is natural to measure expansion directly in terms of eigenvalues themselves. We will concern ourselves with this notion of expansion, called {\it spectral expansion}. 

As smaller values of $\lambda(G)$ mean better expansion properties, it is natural to ask: what is the theoretical minimum of $\lambda(G)$? The Alon-Boppana theorem \cite{Alon1986} answers this question, stating that for a $k$-regular graph with second largest (in magnitude) adjacency eigenvalue $\lambda$ and diameter $D$, we have
$\lambda(G) \geq 2\sqrt{k-1}\left(1-\nicefrac{2}{D}\right)-\nicefrac{2}{D}.$
Consequently, if $(G_i)_{i=1}^{\infty}$ is a family of connected, $k$-regular, $n$-vertex graphs with $n \to \infty$ as $i \to \infty$, then,
$\liminf_{i \to \infty} \lambda(G_i) \geq 2 \sqrt{k-1}.$
We call a graph Ramanujan if it achieves this theoretical minimum, i.e., is an optimal spectral expander. 
\begin{definition} 
A $k$-regular graph $G$ is called Ramanujan if 
$\lambda(G) \leq 2 \sqrt{k-1},$
where $\lambda(G)$ denotes the largest magnitude adjacency eigenvalue of $G$ not equal to $\pm k$. 
\end{definition}

As a consequence of their optimal spectral expansion, Ramanujan graphs possess a plethora of beneficial structural properties discussed in \cite{aksoy2020ramanujan}. In particular, the Ramanujan property guarantees at least nearly optimal bisection bandwidth. 

\begin{figure}
    \centering
\hfill
\begin{subfigure}[t]{.45\columnwidth}
            \includegraphics[trim = 35 35 35 15 ,clip,width=.9\columnwidth]{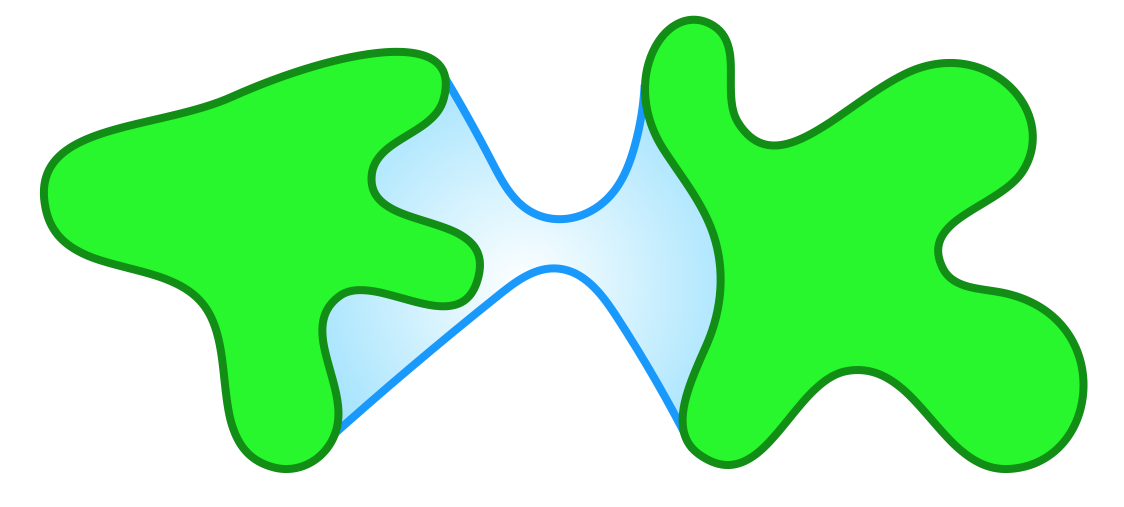}\label{fig:BW_cartoon}%
        \caption{Low Bisection Bandwidth}
   \end{subfigure}
\hskip1.5em
\begin{subfigure}[t]{.45\columnwidth}
        \includegraphics[trim = 60 35 60 60, clip,width=.9\columnwidth]{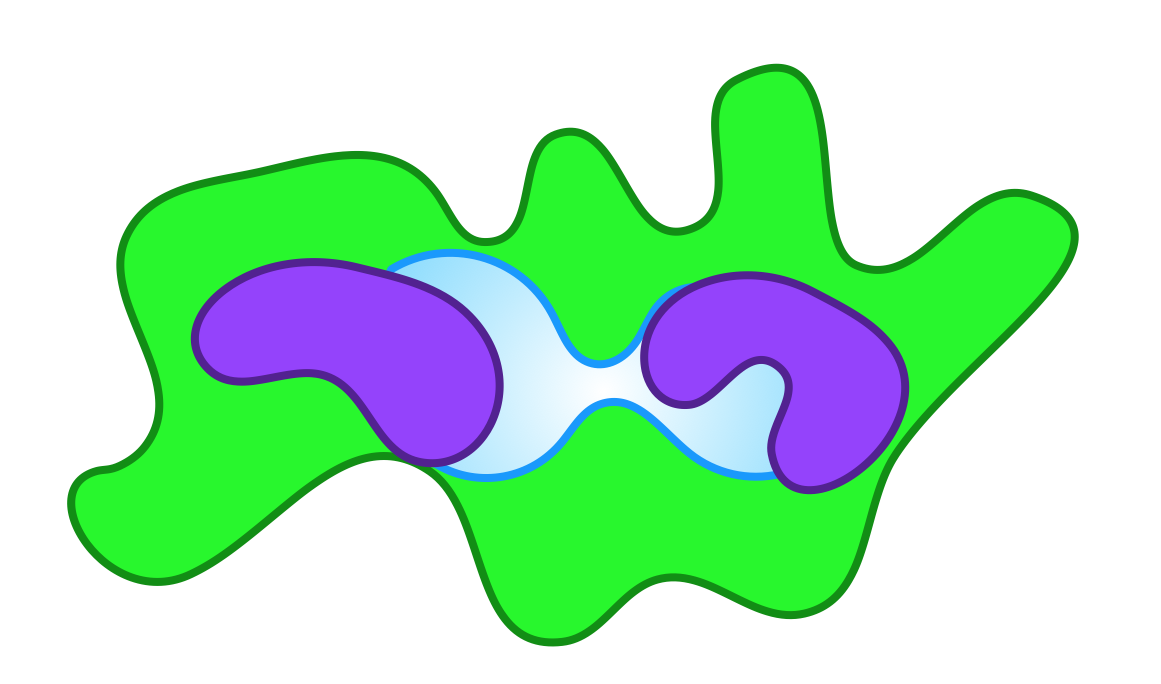}%
        \caption{Low Discrepancy}
    \end{subfigure}
    \hfill \phantom{}
    \caption{Structures forbidden by high bisection bandwidth and discrepancy. Bisection bandwidth only concerns edges crossing a bipartition (blue shadow), while discrepancy also requires any two subsets (in purple), are bottleneck-free. \vspace{-1.5em}}
    \label{F:disc}
\end{figure}

While we emphasize this near-optimality of bisection bandwidth in this work, we note the Ramanujan property guarantees something much stronger: it bounds the number of edges between {\it any} collection of vertices, not just bisections. This stronger property is called the discrepancy inequality~\cite{Chung:spectral}.
Simply put, this means large spectral gap implies two arbitrary subsets of the network are bottleneck-free; see Fig. \ref{F:disc} for a cartoon of substructures forbidden by the discrepancy property.

While we will not explicitly design the experiments of Section \ref{sec:sim} to emphasize the impact of the discrepancy property,
in practice, this is likely to be an important property for the practical usage of the systems.  In particular, as the discrepancy property assures that given an arbitrary collection of vertices involved in a computation the bisection bandwidth on the topology restricted to those vertices is still high, we expect systems designed around Ramanujan graph topologies will be less susceptible to performance degradation based on job schedule and inter-job contention as illustrated in \cite{Bhatele:Dragonfly}.  Additionally, we note that the discrepancy property will likely mitigate the benefit of routing strategies such as Valiant that attempt to homogenize traffic across the network.  In particular, as high discrepancy networks are optimally bottleneck-free, this minimizes the advantage of homogenizing network traffic. 

\paragraph{Related work in HPC} As evidenced by the relationships between eigenvalues and other structural properties highlighted above, it is unsurprising some HPC topologies consider spectral expansion {\it implicitly} in their network design. The well-known randomized Jellyfish topology has strong, albeit not optimal, spectral expansion properties. However, random $k$-regular graphs are ``sub Ramanujan" as shown by Friedman's proof \cite{Friedman2003} of Alon's second eigenvalue conjecture \cite{Alon1986}; hence SpectralFly has superior spectral expansion over JellyFish. Further, as discussed in \cite{valadarsky2015xpander}, the unstructuredness of randomized constructions such as Jellyfish  makes ``them hard to reason about (predict, diagnose), build (e.g., in terms of wiring
complexity), and so poses serious, arguably insurmountable,
obstacles to their adoption in practice." 

Next, we briefly mention other work {\it explicitly} considering notions of spectral expansion or Ramanujan graphs in an HPC setting. 
In \cite{aksoy2020ramanujan}, the authors survey a wide swath of supercomputing topologies and derive either asymptotic bounds or exact expressions for their spectral gap, which shows many supercomputing topologies are far from Ramanujan. In this work, we aim to realize the theoretical potential suggested in \cite{aksoy2020ramanujan} through SpectralFly, which has optimal spectral gap. 
So called $(\alpha,\beta,n,d)-$expanders are utilized to construct ``multibutterfly" networks \cite{Upfal1992}, and later, ``metabutterfly" networks \cite{Brewer1994}, which aim to mitigate wiring complexity.
Valadarsky et al \cite{Valadarsky2016} propose ``Xpander", a general construction in the context of datacenter design. Xpander is based on the theory of graph lifts \cite{Bilu2006} which, via derandomization procedures, may generate deterministic almost-Ramanujan graphs. Theoretical work by Marcus, Spielman and Srivastava \cite{Marcus2013} suggests it may be possible to explicitly generate Ramanujan graphs utilizing $k$-lifts via sophisticated interlacing polynomial techniques. 

\section{\large SpectralFly Topology Construction}\label{sec:const}

Constructing explicit families of Ramanujan graphs is an  ongoing topic of research. The first constructions were by Lubotzky, Phillips and Sarnak \cite{lubotzky1988ramanujan}, and independently, by Marguilis \cite{margulis1988explicit}. In 2013 and 2015, Marcus, Spielman and Srivastava \cite{Marcus2013, Marcus2015} gave new constructions of bipartite Ramanujan graphs. For more on these constructions, see \cite{aksoy2020ramanujan}. 

Here, we focus on the construction by Lubotzky, Phillips and Sarnak, which we refer to as {\it LPS graphs}. These are the graph topologies underlying a SpectralFly network. Hence, when studying graph-theoretic properties, we use the terms ``SpectralFly" and ``LPS" interchangeably. \change{When interpreted as a network, a vertex of an LPS graph corresponds to a router, and edges between vertices correspond to bidirectional links. While fully-realized SpectralFly networks must also specify endpoint concentration (see Section \ref{sec:sim}), in this section we focus on the core LPS topology formed by the routers.

} 

We utilize an extension of the original LPS graphs provided by Morgenstern \cite{Morgenstern1994}. LPS graphs are examples of graphs encoding algebraic group structure, called Cayley graphs.

\begin{definition}[Cayley Graph]
The Cayley graph, $\mathrm{Cay}(\mathcal{G},S)$, of a group $\mathcal{G}$ and symmetric
is a graph on vertex set $V=\mathcal{G}$ and edge set $E=\{\{u,v\}: u^{-1}v \in S\}$.
\end{definition}

An LPS graph is a particular Cayley graph where both the group and the generating set $S$ depend on number-theoretic properties of two input values, $p$ and $q$, as defined below. 

\begin{definition}[LPS Graphs] \label{def:LPS}
The LPS graph $\mbox{LPS}(p,q)$ is a {Cayley graph defined for distinct odd primes $p, q$.}
{To define the generator set,} let $x,y$ be solutions to $x^2+y^2+1=0 \pmod{q}$, and $D$ be the set of solutions $(\alpha_0,\alpha_1,\alpha_2,\alpha_3)$ to $\alpha_0^2 + \alpha_1^2 + \alpha_2^2+\alpha_3^2 = p$ which satisfy
\begin{itemize}
    \item $\alpha_0>0$ is odd, if $p \equiv 1 \pmod{4}$
    \item $\alpha_0>0$ is even, or $\alpha_0=0$ and $\alpha_1>0$, if $p \equiv 3 \pmod{4}$.
\end{itemize}
The generating set $S$ of $\mbox{LPS}(p,q)$ consists of all matrices
\[
 \begin{bmatrix} \alpha_0+\alpha_1x+\alpha_3y & -\alpha_1y + \alpha_2+\alpha_3x \\
-\alpha_1y - \alpha_2 + \alpha_3x & \alpha_0-\alpha_1x-\alpha_3y
 \end{bmatrix},
\]
where $(\alpha_0,\alpha_1,\alpha_2,\alpha_3) \in D$.
The group $G$ of $\mbox{LPS}(p,q)$ is
\[
G=
\begin{cases}
\mathrm{PSL}(2,\mathbb{F}_q) & \mbox{ if } \left(\frac{p}{q}\right)=1 \vspace{1mm}  \\ 

\mathrm{PGL}(2,\mathbb{F}_q) & \mbox{ if } \left(\frac{p}{q}\right)=-1
\end{cases},
\]
where $\left(\tfrac{p}{q}\right)$ is the Legendre symbol, and PSL and PGL 
are the projective special and linear groups, respectively. { If $q > 2\sqrt{p}$, then $\mbox{LPS}(p,q)$ is a $(p+1)$-regular Ramanujan graph.}
\end{definition}

\begin{figure*}[ht]
  \centering
  \phantom{}\hfill
  \includegraphics[width=0.48\linewidth]{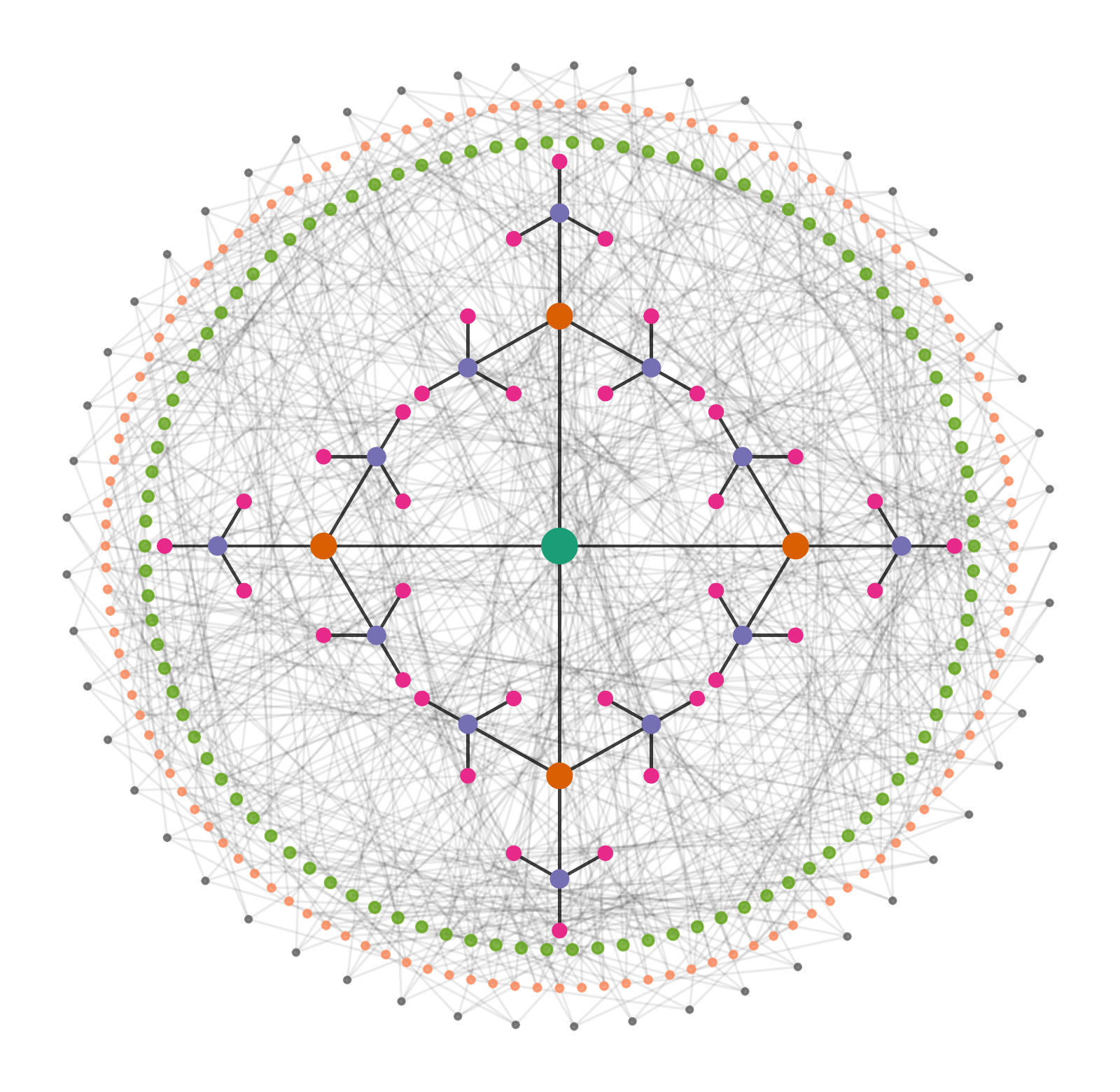} \hfill
  \includegraphics[width=0.48\linewidth]{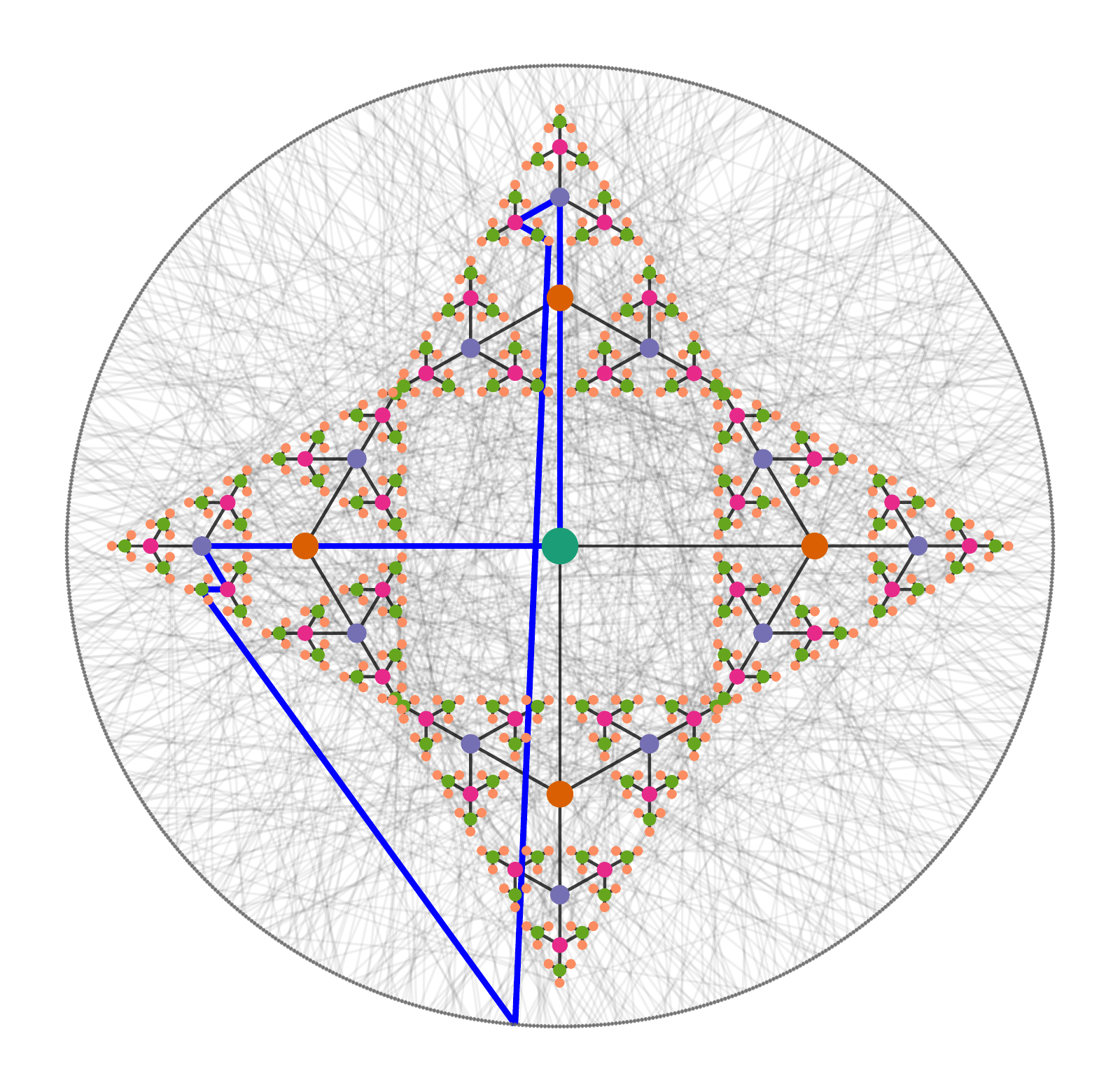} \hfill \phantom{} \\
  \includegraphics[width=0.5\linewidth]{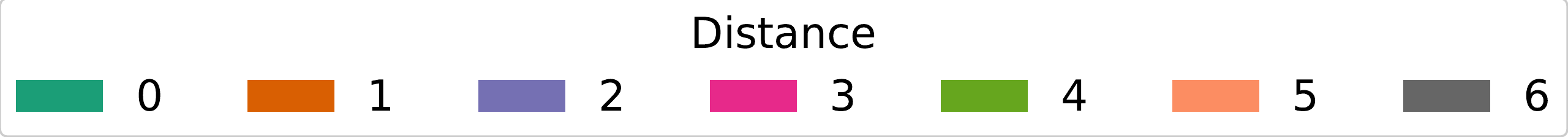}
 \caption{
  Visualization of SpectralFly topology instances: the entire LPS$(3,7)$ graph (left) and the 6-hop neighborhood of a vertex in LPS$(3,17)$ (right). Since LPS graphs are vertex transitive, the $k$-hop neighborhood of every vertex has the same structure. Furthermore, the local neighborhood surrounding a vertex is a tree of variable depth depending on the inputs $p,q$. For instance, a shortest length cycle in LPS$(3,17)$ is highlighted as blue, and utilizes vertices at distance 6 from the center vertex. \vspace{-1.5em} 
 } \label{fig:viz}
\end{figure*}

\change{For those unfamiliar with algebraic graph theory or number theory, a full understanding of the details within the proceeding definition is unnecessary for the purposes of this work (see  \cite{davidoff2003elementary}).
Nonetheless, we include this definition as a self-contained description of the LPS topology. Similarly for completeness and to help garner understanding, we briefly illustrate how to generate LPS graphs in practice with an example below, and include a visualization in Figure \ref{fig:viz}. }
For a full tutorial on LPS graph generation, see \cite{elzinga2010producing}.

\begin{figure}[t] 
    \centering
    \begin{tikzpicture}[square/.style={regular polygon,regular polygon sides=4}]
        \def\x{.4\columnwidth}
        \def\y{1}
        \tikzstyle{element}=[square,draw=black,line width=1pt,align=center, inner sep = 0pt]
        \tikzstyle{action}=[arrows={->[line width=1pt,length=2mm,width=3mm]},line width = 1pt]
        \draw (0,0) node[element, fill=blue!15] (m1) {$\begin{matrix} 0 & 1 \\ 1 & 2\end{matrix}$};
        \draw (\x,\y) node[element, fill=red!15] (m2) {$\begin{matrix}1& 2\\0& 2\end{matrix}$};
        \draw (-\x,\y) node[element, fill=red!15] (m3) {$\begin{matrix}1& 3\\4& 4\end{matrix}$};
        \draw (-\x,-\y) node[element, fill=red!15] (m4) {$\begin{matrix}1& 4\\3& 0\end{matrix}$};
        \draw (\x,-\y) node[element, fill=red!15] (m5) {$\begin{matrix}1& 1\\1& 4\end{matrix}$};
        \draw[action] (m1) -- (m2) node[midway, above, sloped] {\small{$\begin{pmatrix} 1&1\\2&4\end{pmatrix}$}};
        \draw[action] (m1) -- (m3) node[midway, above, sloped] {\small{$\begin{pmatrix} 1&4\\3&4\end{pmatrix}$}};
        \draw[action] (m1) -- (m4) node[midway, below, sloped] {\small{$\begin{pmatrix} 1&2\\1&4\end{pmatrix}$}};
        \draw[action]  (m1) -- (m5) node[midway, below, sloped] {\small{$\begin{pmatrix} 1&3\\4&4\end{pmatrix}$}};
    \end{tikzpicture}
    \caption{ Neighborhood of a vertex in LPS$(3,5)$. Vertices are from $\mathrm{PGL}(2,\mathbb{F}_5)$ labeled by a representative matrix from the coset; edges $\{u,v\}$ are labeled by a generating element $u^{-1}v$ \vspace{-2.5em}}
    \label{F:xwing}
\end{figure}
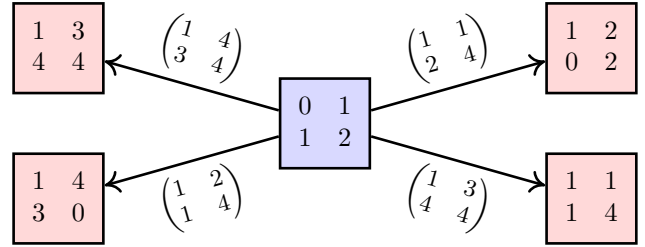

\begin{example}
Let $(p,q)=(3,5)$. These are valid inputs for an LPS graph because $p,q$ are distinct, odd primes and $5>2\sqrt{3}$. Since $x^2 \not \equiv 3 \pmod{5}$ for any $x$, the Legendre symbol $\left( \tfrac{3}{5} \right)=-1$ and hence the group is $\mathrm{PGL}(2,\mathbb{F}_5)$ where $\mathbb{F}_5=\{0,1,\dots,4\}$. The elements of $\mathrm{PGL}(2,\mathbb{F}_5)$ are cosets of $2 \times 2$ matrices with elements in $\mathbb{F}_5$ and nonzero determinant such that $A,B$ are in the same coset if $A=xB$ for some nonzero $x$. For example, the element
\[
v=
\left\{
\begin{bmatrix}
0 & 1 \\
1 & 2
\end{bmatrix},
\begin{bmatrix}
0 & 2 \\
2 & 4
\end{bmatrix},
\begin{bmatrix}
0 & 3 \\
3 & 1
\end{bmatrix},
\begin{bmatrix}
0 & 4 \\
4 & 3
\end{bmatrix}
\right\}
\]
represents a {\it single} element of $\mathrm{PGL}(2,\mathbb{F}_5)$, and hence a single vertex of the graph \mbox{LPS}$(3,5)$. 

Next, we construct the generating set $S$. As $p \equiv 3 \pmod{4}$, we are interested in solutions of $\alpha_0^2 + \alpha_1^2 + \alpha_2^2+\alpha_3^2 = 3$ where either $\alpha_0>0$ and is even, or $\alpha_0=0$ and $\alpha_1>0$. There are no solutions of the former type; solutions of latter type are:
\[
(0,1,1,1),(0,1,-1,-1),(0,1,-1,1),(0,1,1,-1).
\]
Finally, using $(x,y)=(0,2)$ as a solution to $x^2+y^2+1=0 \pmod{5}$, the elements of the generating set $S$ may be constructed. For example, \change{the coset} for the generator $s \in S$ corresponding to the solution $(0,1,1,1)$ is
\[
\left\{
\begin{bmatrix}
1 & 2 \\
1 & 4
\end{bmatrix},
\begin{bmatrix}
2 & 4 \\
2 & 3
\end{bmatrix},
\begin{bmatrix}
3 & 1 \\
3 & 2
\end{bmatrix},
\begin{bmatrix}
4 & 3 \\
4 & 1
\end{bmatrix}
\right\}.
\]
$\mbox{LPS}(3,5)$ is then constructed by creating edges between $u,v \in \mathrm{PGL}(2,\mathbb{F}_5)$ whenever $us=v$ for $s \in S$. Figure \ref{F:xwing} illustrates the neighborhood of a certain vertex in LPS$(3,5)$, labeling edges by the associated element $s \in S$.
\end{example}

\begin{figure}[ht!]
    \centering
    \includegraphics[width=0.49\columnwidth]{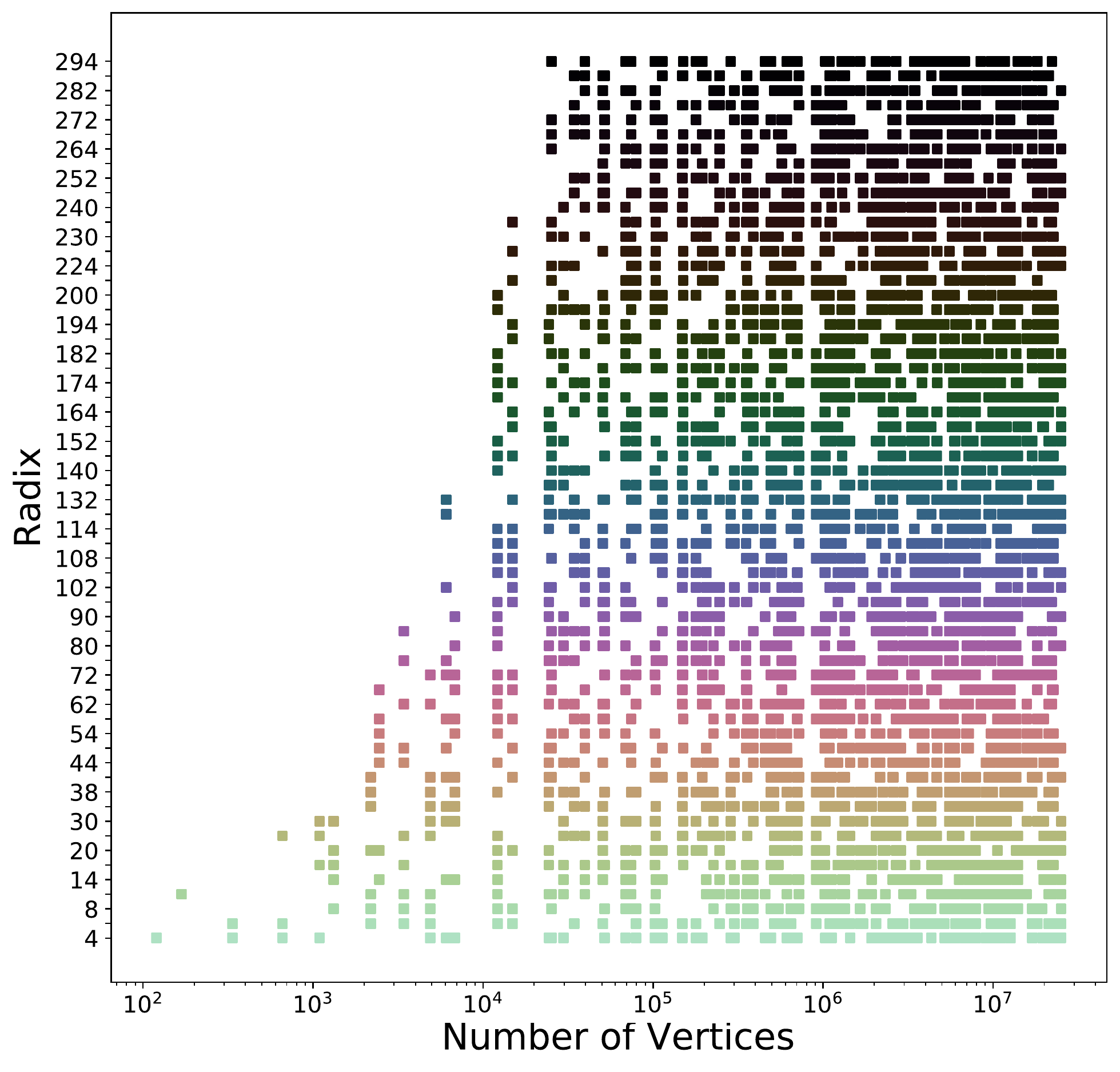}
    \includegraphics[width=0.49\columnwidth]{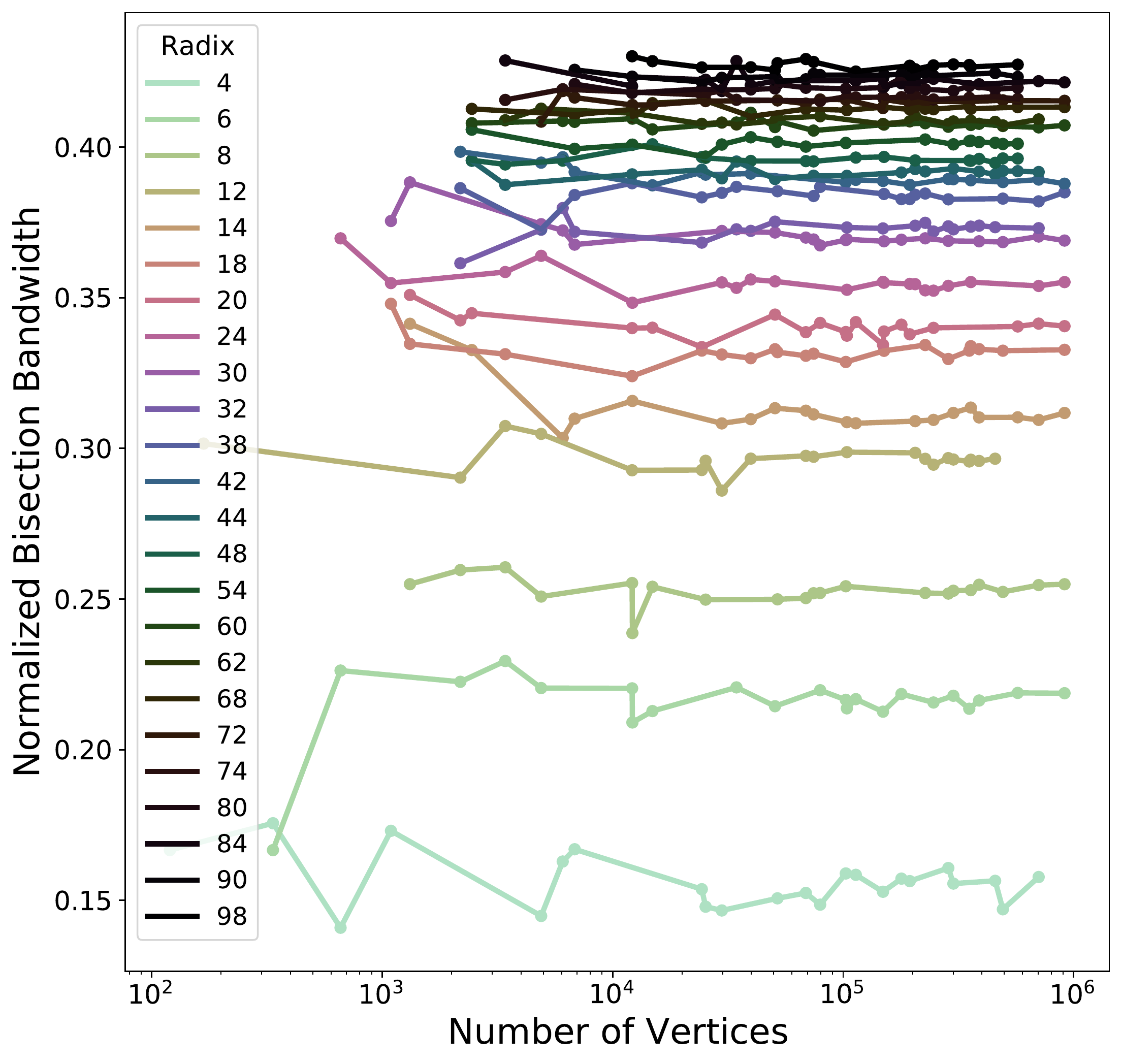} \\
      \includegraphics[clip,width=0.49\columnwidth]{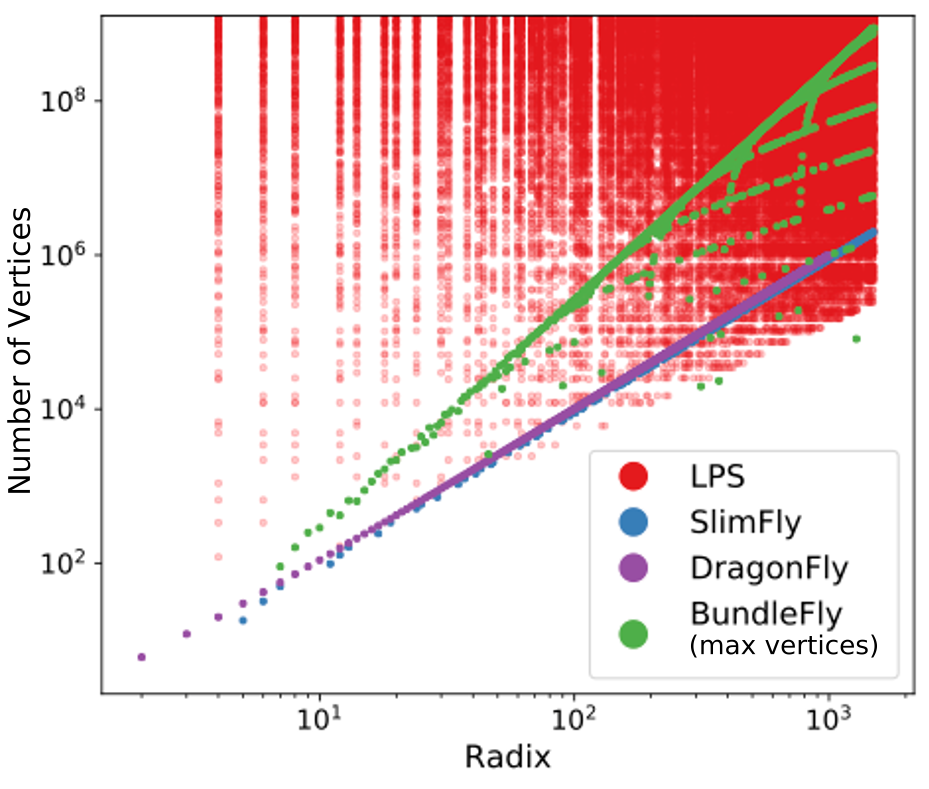}\label{fig:sizeComp}%
        \includegraphics[clip,width=0.49\columnwidth]{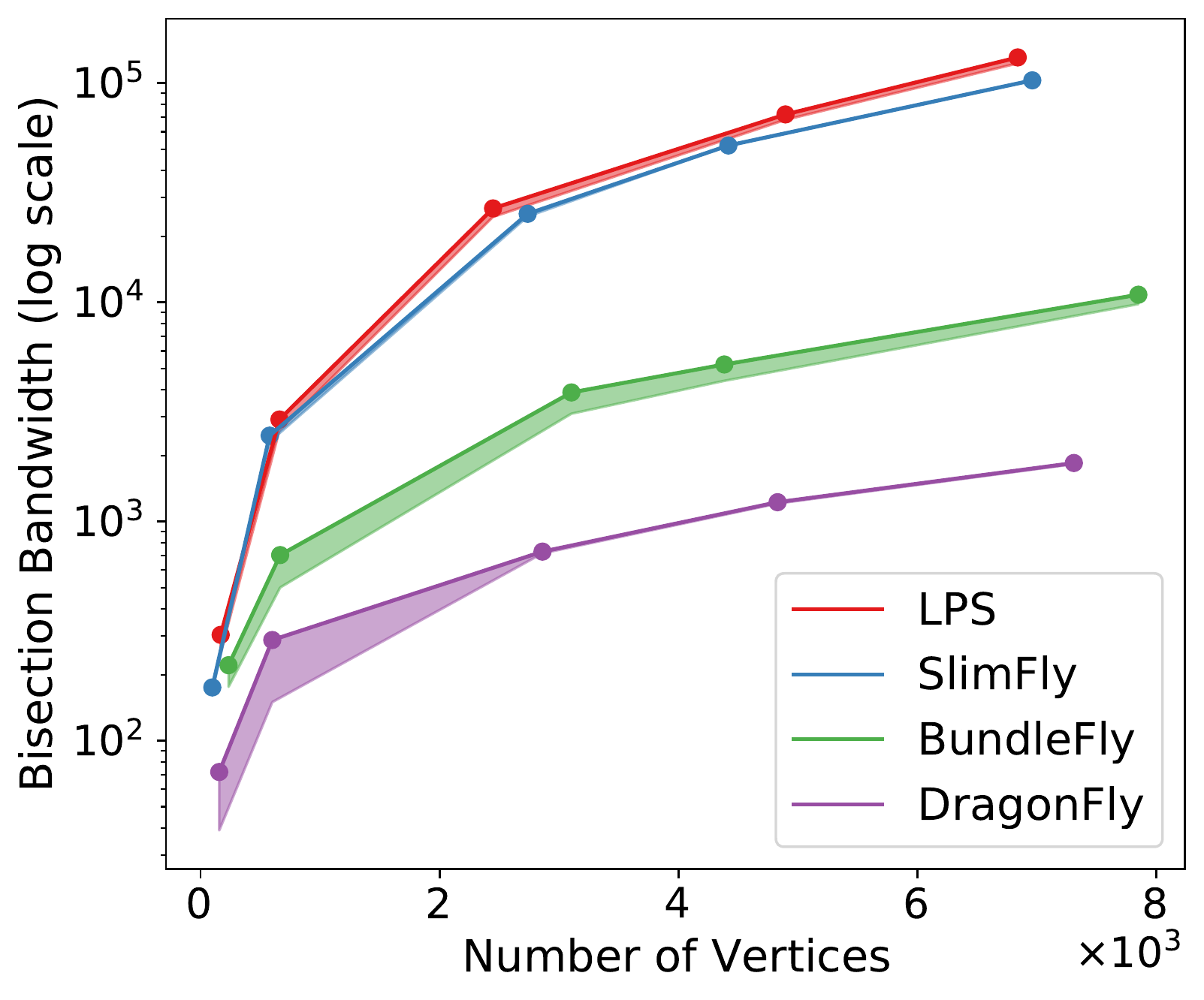}%
    \vspace{-0.5em}
    \caption{Possible number of vertices and radix of LPS for $p,q<300$ (upper left), normalized bisection bandwidth of LPS for $p,q<100$ (upper right), feasible topology sizes per radix (lower left), raw bisection bandwidth comparison (lower right). \vspace{-3em}}
    \label{fig:sizes_bws}
\vspace{-0.2em}    
\end{figure}

There are several reasons for selecting LPS graphs amongst currently proposed Ramanujan constructions. First, LPS graphs are flexible in terms of feasible sizes and radix values. Figure \ref{fig:sizes_bws} (left) plots radix and vertex counts for all possible LPS graphs generated with inputs $p,q<300$. While, like almost any structured family of topologies, some radix and vertex count combinations are infeasible, the absence of large gaps in the plot suggests the high likelihood of finding an LPS graph ``acceptably close" to any given desired radix and vertex count combination. As discussed further in Section \ref{sec:structProp}, this flexibility stands in contrast to many competing graph topologies. LPS graphs afford users the ability to generate {\it arbitrarily large} graphs for a given radix, whereas the sizes of other topologies can only be increased via the radix.  

Secondly, in addition to exhibiting the Ramanujan property, LPS graphs possess other desirable characteristics.
For example, since LPS graphs are Cayley graphs, they are vertex-transitive. Informally, this means every vertex has an identical local environment, i.e. the graph ``looks the same" from every vertex. Thus, the 6 hop neighborhood of a vertex in LPS$(3,17)$ seen in Figure $\ref{fig:viz}$ has an identical structure for all vertices. Consequently, vertex-transitivity enables simplifications which benefit the computational cost and design of routing protocols. 
Their algebraic structure also affords other benefits, such as optimal edge-connectivity (a key consideration for network resiliency) as well as efficient algorithms by which topologies on tens of millions of vertices may be easily generated \cite{elzinga2010producing}.

In addition to possessing these properties by virtue of being a Cayley graph, LPS graphs are also widely studied. Over the past several decades, researchers have bounded or characterized the diameter, path length behavior, and fault tolerance of LPS graph, making them attractive options for supercomputing topologies, as argued in \cite{aksoy2020ramanujan}. 
One such key property for interconnection networks is bisection bandwidth. Figure \ref{fig:sizes_bws} (upper right) presents the normalized bisection bandwidth of LPS graphs for various sized topologies on radix values between $k=4$ and 98, divided by $nk/2$ to ensure a size-agnostic comparison.  We observe larger normalized bisection bandwidth values are achieved for larger radix graphs, with diminishing returns. In contrast to some other topologies we survey, the bisection bandwidth doen't decay as LPS graph size increases per fixed radix, which is a consequence of the Ramanujan property. Furthermore, larger normalized bisection bandwidth values are feasible for larger radix networks.

\section{Structural Property Comparison}\label{sec:structProp}

\begin{table}
\footnotesize
\begin{tabular}{l|c|c|c|c|c|c}
    \multirow{2}{*}{\bf Topology} & \multirow{2}{*}{\bf \change{Routers}} &{\bf \change{Router}} & \multirow{2}{*}{\bf Diam.} & \multirow{2}{*}{\bf Dist.} & \multirow{2}{*}{\bf Girth} & \multirow{2}{*}{\boldsymbol{$\mu_1$}}  \\
   &  & {\bf Radix} &  &  &  &   \\
     \toprule
    LPS$(11,7)$ & 168 & 12 & 3 & 2.39 & 3 & 0.50 \\
    SF$(7)$ & 98 & 11 & 2 & 1.89 & 3 & 0.62 \\
    BF$(13,3)$ & 234 & 11 & 3 & 2.56 & 3 & 0.27 \\
    DF$(12)$ & 156 & 12 & 3 & 2.70 & 3 & 0.08 \\
    \hline
    LPS$(23,11)$ & 660 & 24 & 3 & 2.35 & 3 & 0.65 \\
    SF$(17)$ & 578 & 25 & 2 & 1.96 & 3 & 0.64 \\
    BF$(37,3)$ & 666 & 23 & 3 & 2.61 & 3 & 0.13 \\
    DF$(24)$ & 600 & 24 & 3 & 2.84 & 3 & 0.04 \\
    \hline
    LPS$(53,17)$ & 2448 & 54 & 3 & 2.32 & 3 & 0.74 \\
    SF$(37)$ & 2738 & 55 & 2 & 1.98 & 3 & 0.65 \\
    BF$(97,4)$ & 3104 & 54 & 3 & 2.76 & 3 & 0.07 \\
    DF$(53)$ & 2862 & 53 & 3 & 2.93 & 3 & 0.02 \\
    \hline
    LPS$(71,17)$ & 4896 & 72 & 4 & 2.61 & 4 & 0.77 \\
    SF$(47)$ & 4418 & 71 & 2 & 1.98 & 3 & 0.66 \\
    BF$(137,4)$ & 4384 & 74 & 3 & 2.76 & 3 & 0.05 \\
    DF$(69)$ & 4830 & 69 & 3 & 2.94 & 3 & 0.01 \\
    \hline
    LPS$(89,19)$ & 6840 & 90 & 4 & 2.61 & 4 & 0.80 \\
    SF$(59)$ & 6962 & 89 & 2 & 1.99 & 3 & 0.66 \\
    BF$(157,5)$ & 7850 & 85 & 3 & 2.82 & 3 & 0.06 \\
    DF$(85)$ & 7310 & 85 & 3 & 2.95 & 3 & 0.01 \\
    \bottomrule
\end{tabular}
\caption{Basic structural properties \vspace{-1.5em}}\label{tab:structProp}
\vspace{-1em}
\end{table}

In order to understand the trade-offs between costs, diameter, and bisection bandwidth we compare the combinatorial properties of four topologies representing extreme points at or near the design space Pareto frontier.  Specifically, we consider the DragonFly (optimizing cost and diameter), SlimFly (optimizing diameter and size), BundleFly (optimizing diameter and cost), and LPS/SpectralFly (optimizing spectral gap).

Since random graph constructions, such as the aforementioned JellyFish, have sub-optimal spectral gap \cite{Friedman2003}, and also face serious challenges to adoption in practice due to their unstructuredness, we limit our comparison to {\it deterministic} topologies. Furthermore, we've selected topologies capable of being scaled to beyond tens of thousands of vertices, and which are flexible enough to generate instances with similar size, radix and link counts to other topologies, in order to ensure a fair comparison. Satisfying these criteria, the topologies we consider are defined as follows:

\begin{itemize}[noitemsep,topsep=0pt]
    \item LPS$(p,q)$: The topology underlying SpectralFly, LPS graphs \cite{lubotzky1988ramanujan} are described in Definition \ref{def:LPS}. The radix is $p+1$ and the number of vertices is 
    \change{$\left(3-\left(\tfrac{p}{q}\right)\right)\left(\nicefrac{q^3-q}{4}\right)$}.
    \item SlimFly, SF$(q)$: Studied in \cite{besta2014slim}, SlimFly topologies are based on the MMS graph construction by McKay, Miller and \v{S}ir\'{a}\v{n} \cite{mckay1998note}. For a description of the MMS graph construction, see \cite{hafner2004geometric}. The number of vertices is $2q^2$ and radix is $\frac{3q-\delta}{2}$, where $q=4k+\delta$ for $\delta \in \{-1,0,1\}$. 
    \item BundleFly, BF$(p,s)$: a multi-star product of an MMS graph with parameter $s$ and Paley graph with parameter $p$ -- see \cite{lei2020bundlefly}. The number of vertices is $2ps^2$ and the radix is $\frac{p-1}{2}+\frac{3s-\delta}{2}$ where $s=4k+\delta$ for $\delta \in \{-1,0,1\}$.
    \item DragonFly, DF$(a)$: while there are many DragonFly variants (see \cite{teh2017design,aksoy2020ramanujan} for specifications), we consider the ``canonical" DragonFly topology consisting of $a+1$ fully connected groups, each on $a$ vertices. The number of vertices is $a(a+1)$ and the radix is $a$.
\end{itemize}
While it would be interesting to explore properties of Xpander topologies here, applying complicated interlacing polynomial approaches for their construction and the need to calculate the set of all shortest paths for every pair of routers makes such an evaluation impractical at scales of interest.

We consider 5 size classes for each topology, ranging from ~100 vertices to ~7K vertices. For each size class, we conduct a parameter search to select the topology with closest radix and number of vertices relative to the others in that class. Table \ref{tab:structProp} shows the 4 topologies within each size class have very close radix, and fairly close node counts ensuring a fair comparison \change{of the performance based on the \emph{structural properties} of these networks} in our subsequent experiments. \change{The topologies also have similar {\it girth} (length of the shortest cycle), with larger LPS topologies being the sole examples of girth 4 topologies. }

\paragraph{Feasible topology sizes per radix} The LPS construction accommodates a variety of radix and node size combinations. {In general, Ramanujan graphs of any size are possible; however the smallest possible LPS graph is on 120 vertices}. Fig. \ref{fig:sizes_bws} (lower left) plots possible vertex count and radix combinations. For SlimFly and canonical DragonFly, a large, low-radix topology is impossible, as the radix uniquely determines the topology size. BundleFly allows multiple possible vertex sizes per radix, but the choice of radix constrains the possible vertex sizes. The green points plot the maximum possible number of vertices per each feasible BundleFly radix. Some of maxima drop off sharply for certain radix values, suggesting the range of possible sizes may be unstable.

\paragraph{Diameter and average path lengths}
As summarized in 
Table \ref{tab:structProp}. 
SlimFly always has diameter 2, while BundleFly and DragonFly have diameter 3. In contrast, the diameter of LPS graphs depends on the topology size; numerical experiments from \cite{sardari2019diameter} suggest this diameter is asymptotic to $(4/3)\log_5(n)$. 
LPS has the second smallest average shortest path length (i.e. distance) across all size classes, in spite of sometimes having the largest diameter (for the fourth and fifth size classes). This gap between diameter and average distance suggests ``most'' pairs of vertices in LPS graphs may be closer in distance than the diameter. This is also apparent in Figure \ref{fig:viz}'s visualization of LPS$(3,7)$, where relatively fewer vertices appear at distance equal to the diameter from the center vertex. 
Indeed, recent work by Sardari \cite{sardari2019diameter} proved for any $k$-regular Ramanujan graph, only a tiny fraction of all pairs of vertices have distance greater than $(1+\varepsilon)\log_{k-1}(n)$. Furthermore, for each vertex $x$, the number of vertices at distance greater than this exponentially decays, being less than $n^{1-\varepsilon}$.

\paragraph{Normalized Laplacian spectral gap, $\mu_1$}

To enable cross-size comparison, we compute the normalized Laplacian spectral gap, $\mu_1$, related to the second largest adjacency eigenvalue $\lambda$ by $\mu_1=\nicefrac{k-\lambda}{k}$,
where $k$ is the radix. Whereas smaller values of $\lambda$ ensure better spectral expansion, this is associated with {\it larger} values of $\mu_1$.
Compared with SlimFly and LPS, Table \ref{tab:structProp} shows BundleFly and DragonFly with smaller values of $\mu_1$, which decay for larger sized topologies. 
As proven in \cite{aksoy2020ramanujan}, the second normalized Laplacian eigenvalue of the SlimFly topology SF$(q)$ is $\frac{2}{3+\delta/q}$ \textapprox{}$\frac{2}{3}$. Since LPS graphs are Ramanujan, they have $\mu_1$ at least as large as $\frac{k-2\sqrt{k-1}}{k}$. Thus an LPS graph with radix $k\geq 35$ is guaranteed to have larger $\mu_1$ than {\it any} SlimFly topology. LPS graphs with smaller radix values may still have larger $\mu_1$ (as seen in the second size class in Table \ref{tab:structProp}) or smaller $\mu_1$ (as seen in the first size class). 

\paragraph{Bisection bandwidth}

We use METIS \cite{karypis1997metis} to approximate bisection bandwidth, establishing an upper bound given by the points in Fig. \ref{fig:sizes_bws} (lower right). We also compute a lower bound from \cite{fiedler1973algebraic},
$\mbox{BW}(G)\geq \frac{\lambda_1 k n}{4}$,
where $k$ is the radix and $n$ is the number of vertices. The exact bisection bandwidth lies between these points, represented by the shaded regions.
Recall we are considering the router topology, without regard to a specific concentration. While one can further analyze bisection bandwidth under a particular concentration level, the relative orderings we observe here also hold whenever chosen concentration levels are equal, and so we omit this design choice for clarity and simplicity. 
As seen in Fig. \ref{fig:sizes_bws} in log-scale, 
 as the size of SlimFly topologies increase, the gap between its normalized bisection bandwidth and that of a similar radix LPS widens further. SpectralFly has up to a 39\% increase in bisection bandwidth over SlimFly. 
 This can be confirmed analytically: applying bounds from \cite{aksoy2020ramanujan}, the normalized bisection bandwidth of SlimFly is asymptotically $1/3$. LPS graphs have normalized bisection bandwidth at least $\frac{k-2\sqrt{k-1}}{2k}$, guaranteeing an LPS graph with $k\geq 36$ has larger normalized bandwidth than {\it any} SlimFly.  We emphasize this is a {\it lower bound}; the normalized bisection bandwidth of LPS graphs computed by METIS exceeds $1/3$ around radix 18. 

\subsection{Structural Properties Under Link Failures}

\begin{figure}[]
  \includegraphics[clip,width=0.5\columnwidth]{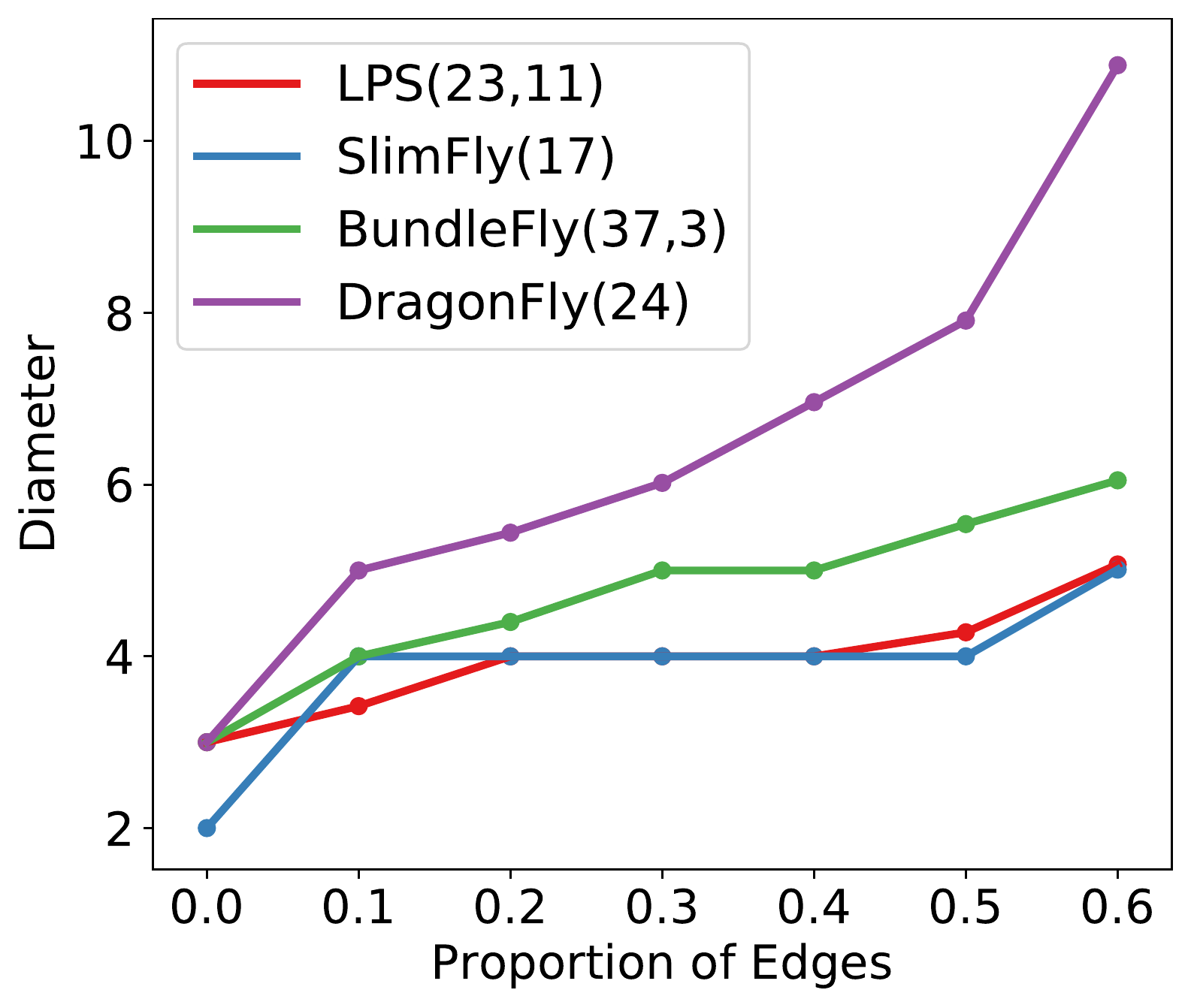}%
  \includegraphics[clip,width=0.5\columnwidth]{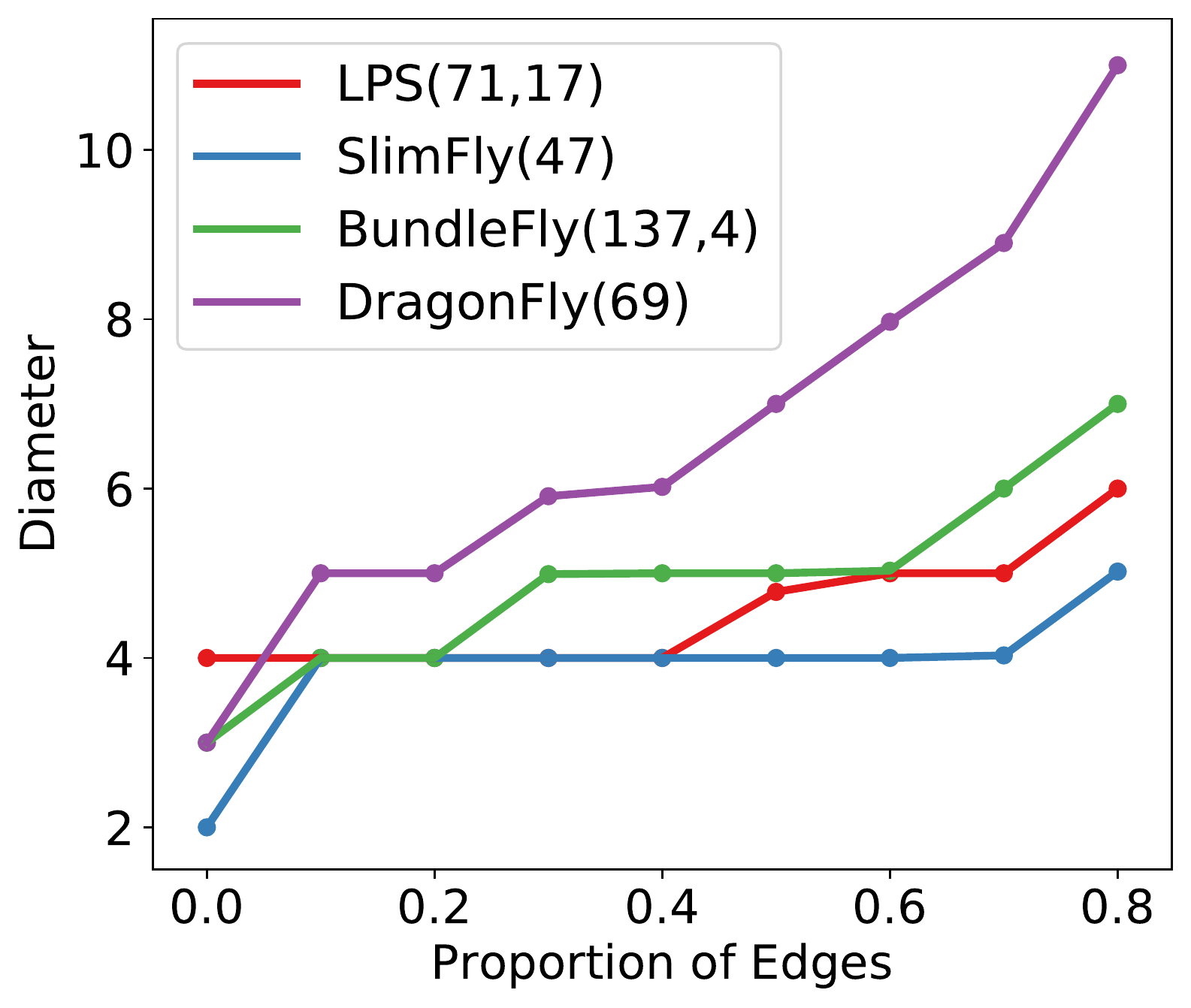} \\
  \includegraphics[clip,width=0.49\columnwidth]{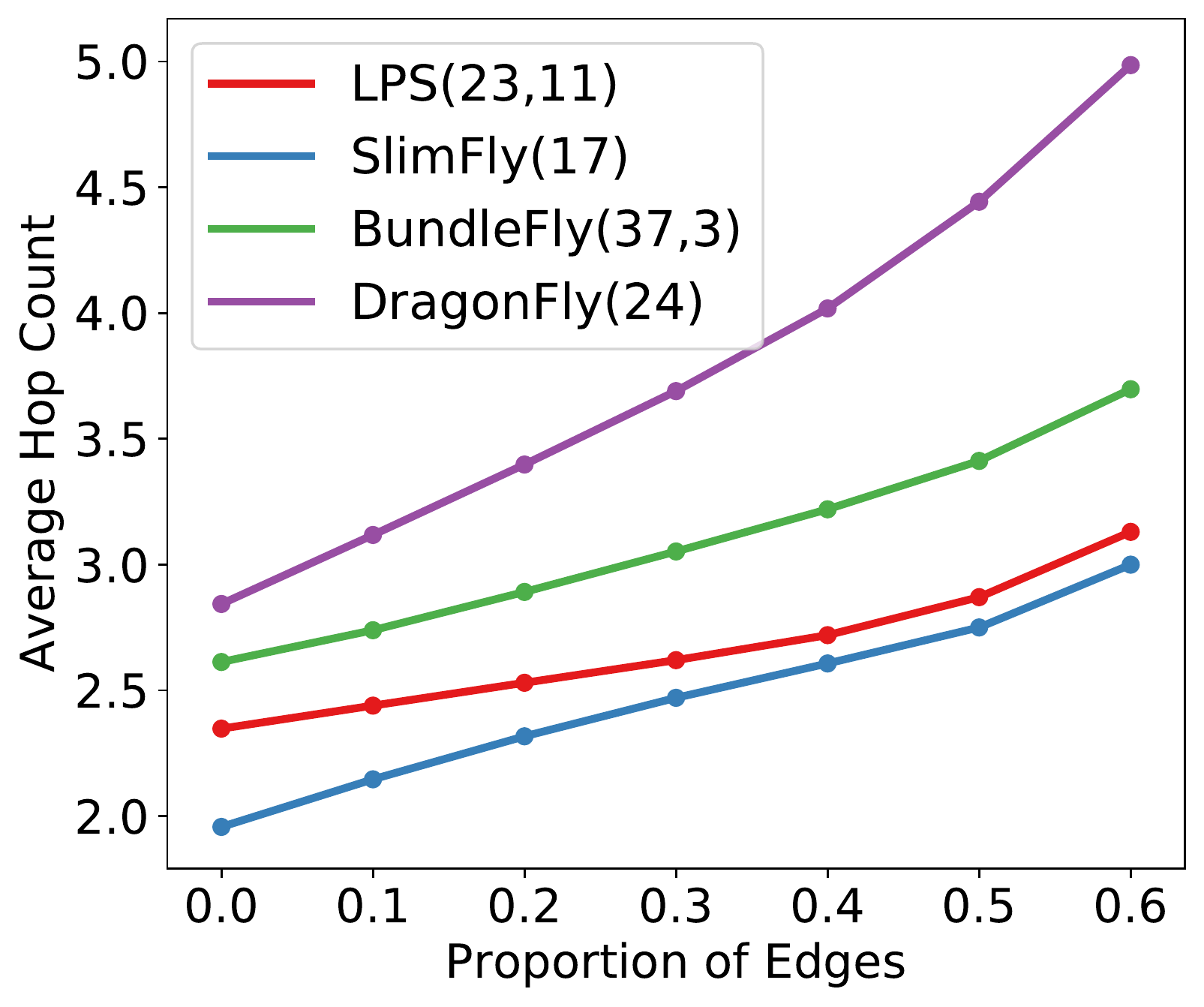}
  \includegraphics[clip,width=0.49\columnwidth]{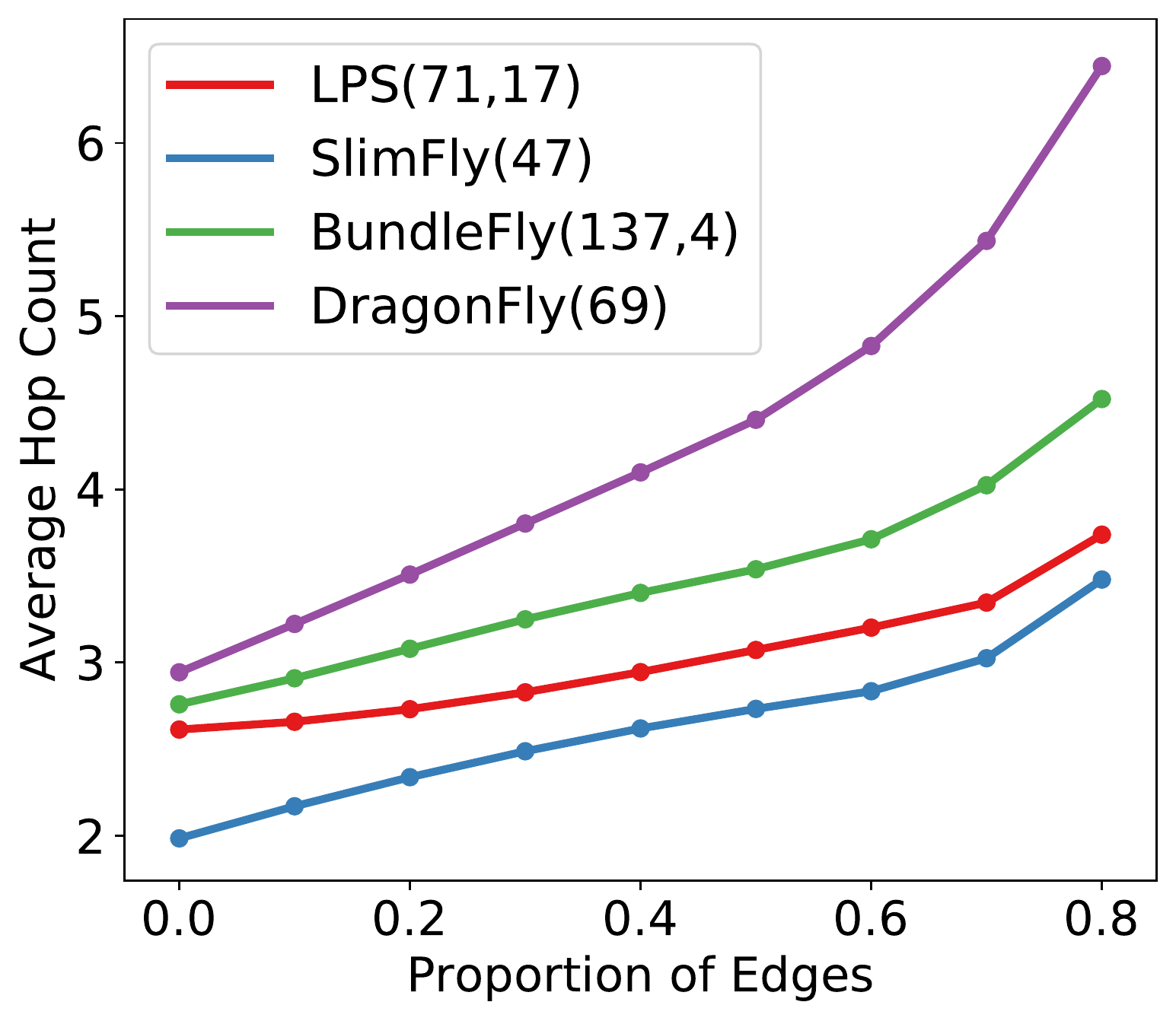} \\
  \includegraphics[clip,width=0.5\columnwidth]{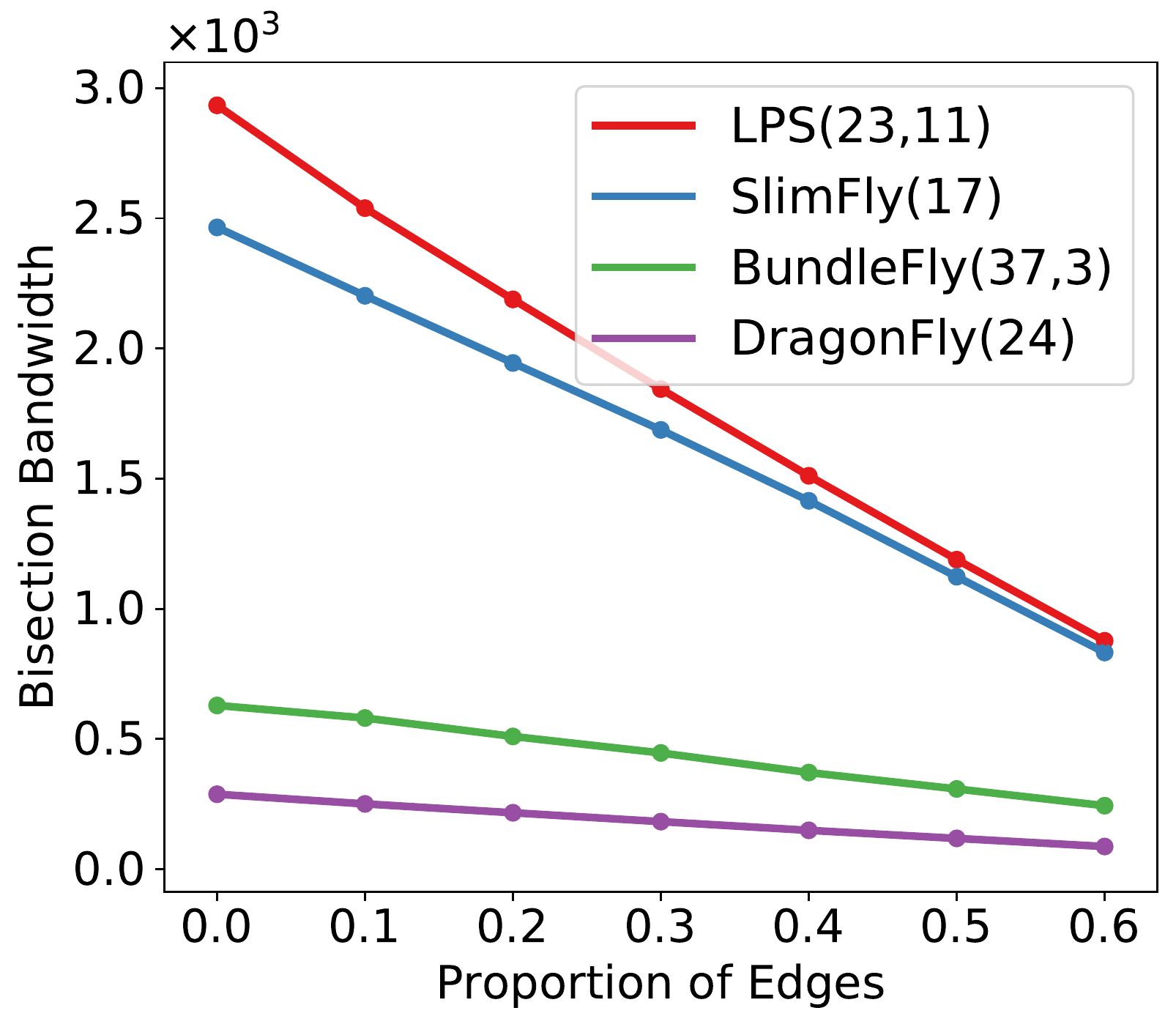}%
  \includegraphics[clip,width=0.49\columnwidth]{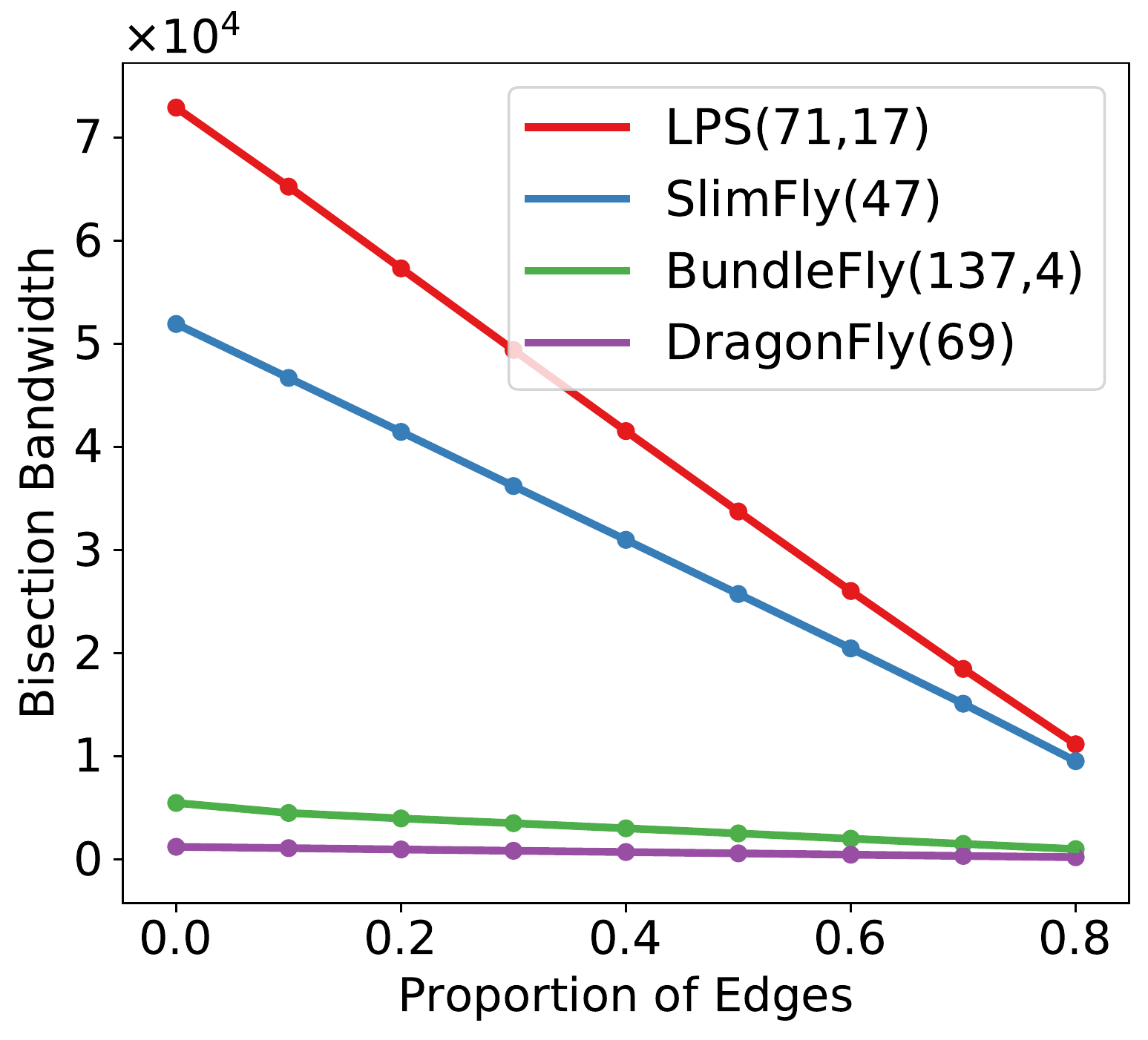}%

\caption{
Structural properties under edge failures for comparable LPS, SlimFly, BundleFly and DragonFly topologies on about 600 vertices (left column) and 7K vertices (right) \vspace{-1.5em}}\label{fig:deleteAll}
\end{figure}

We also examine how these structural properties vary under link failures of varying magnitudes.For each topology, we delete $k$ proportion of its edges, chosen randomly. Our results are averaged over sufficiently many trials.\footnote{For each topology, proportion $k$, and structural property measured, we increase the number of trials $x$ in powers of 10 until the coefficient of variation of sample means across 10 batches of $x$ trials is less than $10\%$.} We run these experiments for ``small", ~600 vertices, instances of each topology, as well as intermediate sized topologies on ~5K vertices. Figure \ref{fig:deleteAll} presents the results for diameter, mean distance, and bisection bandwidth. Note these measures are only well-defined for {\it connected} topologies; however, all four topologies remain consistently connected for small (left column) and medium (right column) sizes under random link failures until 60\% and 80\%, respectively, of edges are deleted. Thus, we only consider edge deletion proportions up until this disconnection threshold. 

With regard to diameter, SlimFly has the smallest value of 2 of the topologies surveyed. However, at 10\% edge failure, this diameter increases to $4$, while LPS topologies exhibit slightly smaller diameter. This suggests SlimFly diameter is more {\it fragile} than that of LPS, congruent with our prior observation that while nearly every pair of vertices in SlimFly is separated by a 2 hop distance, only very few pairs of vertices in an LPS graph achieve the diameter \cite{sardari2019diameter}. 
While LPS maintains a slight edge over SlimFly for 10\% edge failures, for 20-50\% edge failures they have comparable diameter, and for $>50$\% SlimFly has slightly smaller diameter.

Lastly, for mean distance and  bisection bandwidth, LPS and SlimFly perform the best. SlimFly has the smallest mean distance across all edge failure rates, with the gap between LPS narrowing slightly as a higher proportion of edges fail. For bisection bandwidth, LPS retains its larger bandwidth over SlimFly; this gap narrows significantly beyond 20\% failure.

In summary, LPS and SlimFly are consistently more resilient under random edge failures than BundleFly and DragonFly with regard to diameter, average distance, and bisection bandwidth. For diameter, LPS and SlimFly are comparable, with LPS having slightly better diameter for 10\% edge failures and worse for 50\% and above. For average distance and bisection bandwidth, SlimFly retains lower hop count while LPS retains superior bisection bandwidth.

\section{Routing Algorithms}\label{sec:route}
We consider 3 types of routing strategies for SpectralFly: shortest path routing (minimal), Valiant routing, and Universal Global Adaptive (UGAL) routing. In minimal routing, given a source-destination pair $(s,d)$, a packet is forwarded along the routers on the shortest path 
from $s$ to $d$. In theory, minimal routing  will minimize the overall latency of communication thereby outperforming other routing schemes when the underlying network has no congestion. However, in a congested network, shortest paths  may not be the best choice for routing. This is especially true when the betweenness centrality scores of a set of vertices (routers) in a graph (topology) are quite high, meaning these set of vertices will be on the shortest paths for many vertices in the graph. Consequently these vertices will become the bottlenecks in a highly-saturated network.

Congestion concerns have prompted alternative routing schemes which improve performance on various topologies. One such alternative to shortest path routing is Valiant routing~\cite{valiant1982scheme} which proceeds in two phases: given a source-destination pair $(s,d)$, a random intermediate router $i$ is chosen. The packet is then routed from $s$ to $i$ along a shortest path. Once the packet arrives at $i$, the second phase forwards the packet from $i$ to $d$ by following a shortest path.

However, Valiant routing ignores the current state of routers, such as queue length. To ameliorate this, the UGAL family of routing protocols selects dynamically between the minimal path or a Valiant-style paths based on the current state of the system.  For example, in the UGAL-L variant, each router only maintains information about the queue lengths of the local outports.  Using this information at the source, a packet is either forwarded to a random intermediate node first or follows a minimal path based on the queue sizes of the local random outport and minimal outport and total hopcounts from the source to the destination for these two possible routes. 

\subsection{Deadlock Avoidance}
Due to limited resources on each router (buffer count, size, etc.), cyclic dependencies can arise in the resource dependency graph, where messages may try to flow from one router to the next but also messages from the next router may try to flow in the reverse direction. As the buffers fill up and traffic from each router blocks each other in a cycle, this ultimately results in a deadlock.  
Such deadlocks can be avoided primarily in three ways: (1) by creating an acyclic routing scheme; (2) by using virtual channels (VC) and changing the virtual channel to route a packet on each network hop (by incrementing the virtual channel on each network hop, deadlock-free routing can be guaranteed); (3) by running a cycle-detection algorithm on the routing graph beforehand. Each time a cycle is detected, a new virtual channel is added to one of the routing edges, continuing until there are no more cycles. We've chosen the second approach to avoid deadlock based on virtual channels, since it does not require any preprocessing of the topology graph. We set the number of  virtual channels to be equal to the diameter of SpectralFly, $d+1$ for the shortest path routing and $(2d + 1)$ for Valiant routing. 

\begin{figure*}[tp]
\centering
\begin{subfigure}[b]{0.26\textwidth}
    \centering
  \includegraphics[width=\linewidth]{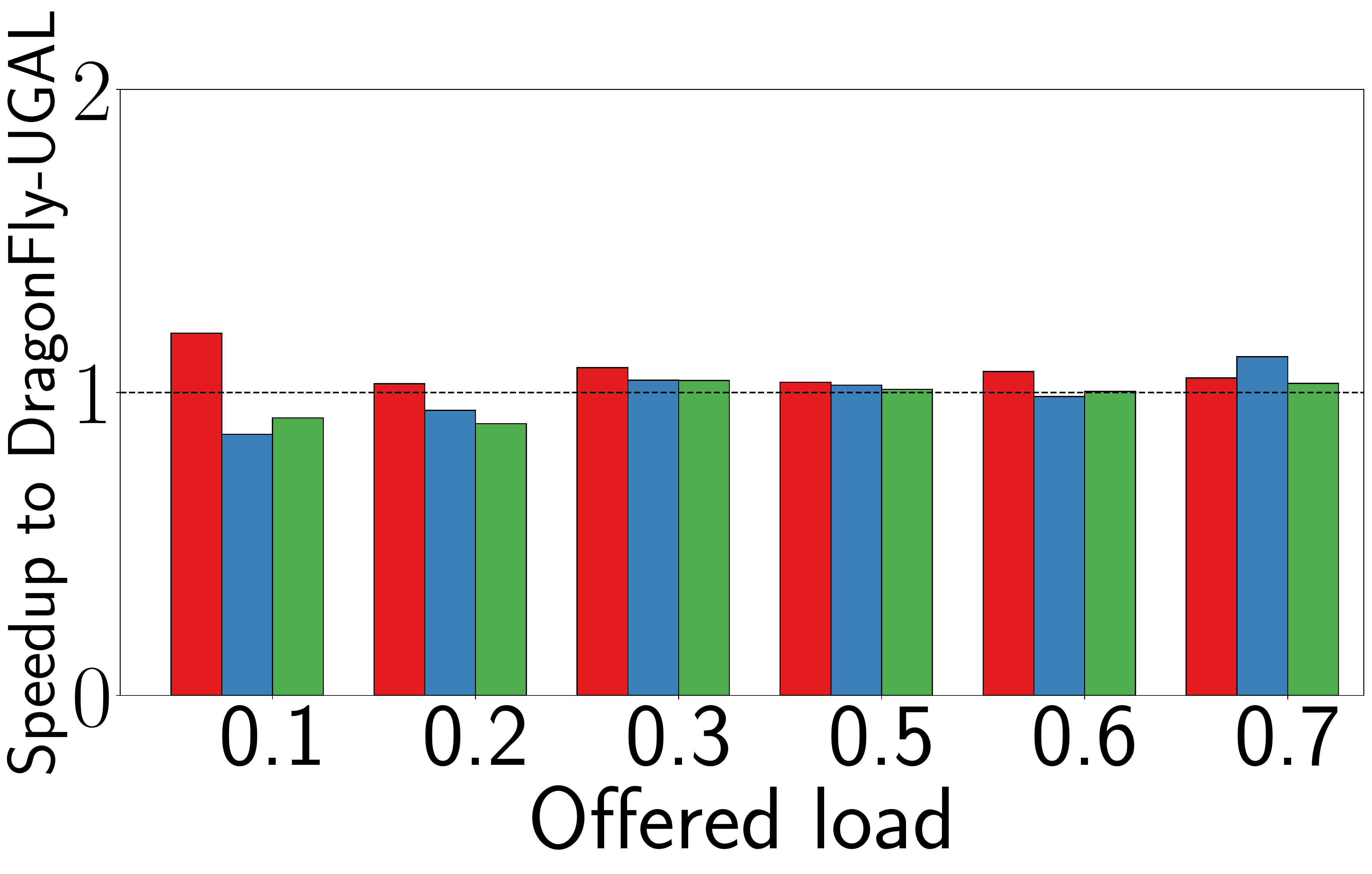}
  \caption{Random.}
  \label{fig:offered_load_random_ugal}
  \end{subfigure}%
\begin{subfigure}[b]{0.25\textwidth}
    \centering
  \includegraphics[width=\linewidth]{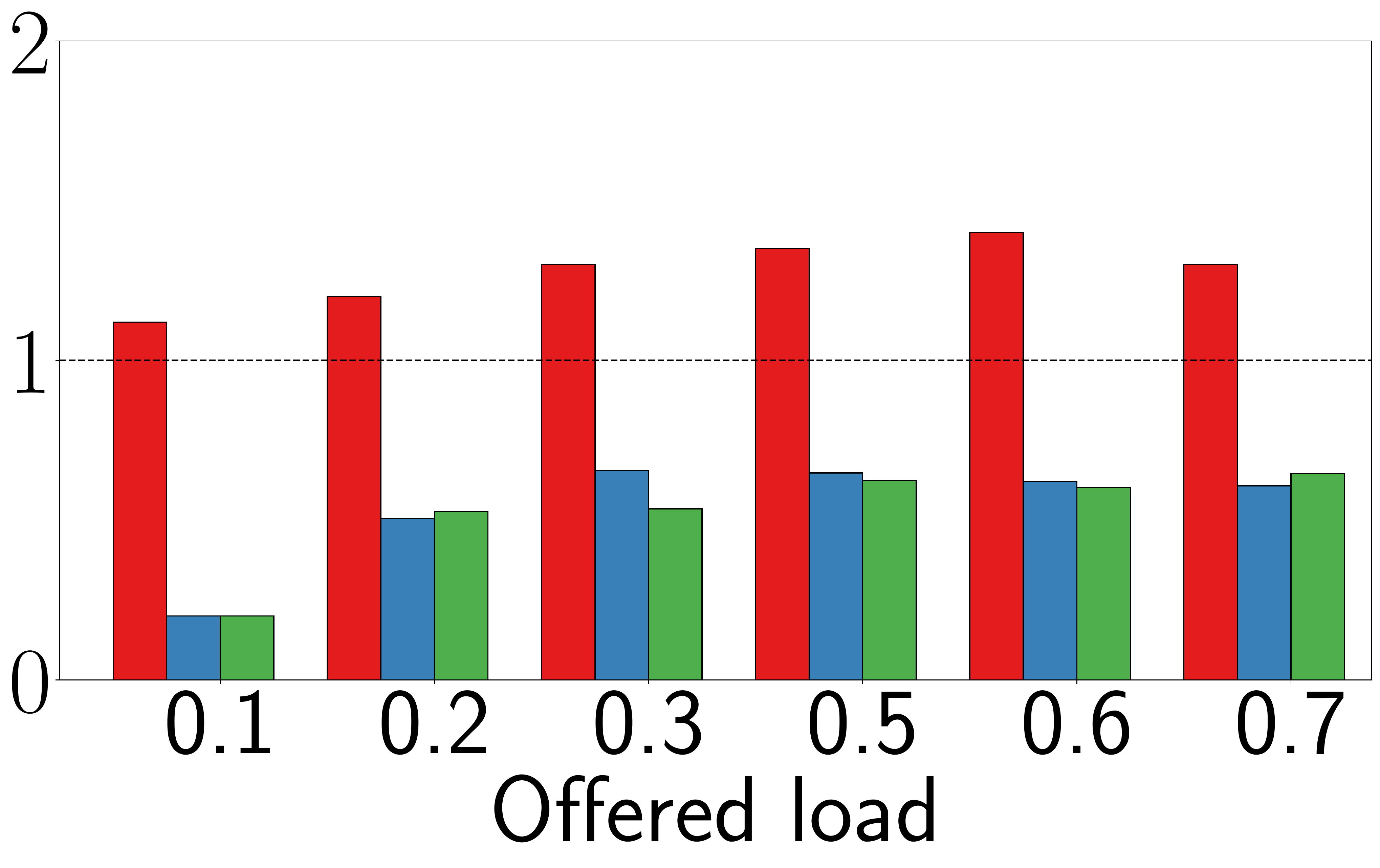}
  \caption{Bit shuffle.}
  \label{fig:offered_load_bitshuffle_ugal} 
  \end{subfigure}%
\begin{subfigure}[b]{0.25\textwidth}
    \centering
  \includegraphics[width=\linewidth]{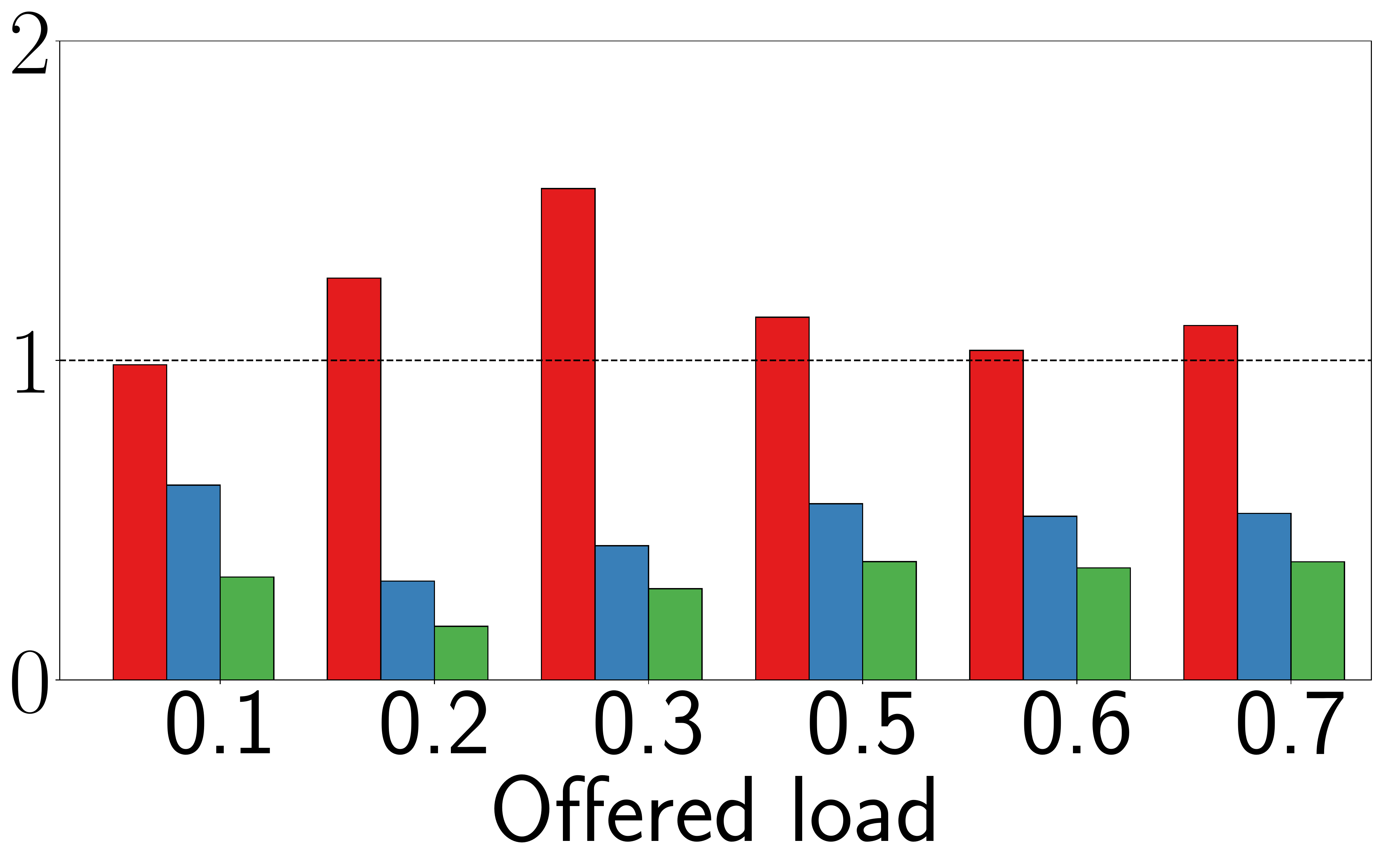}
  \caption{Bit reverse.}
  \label{fig:offered_load_bitreverse_ugal} 
  \end{subfigure}%
\begin{subfigure}[b]{0.25\textwidth}
    \centering
  \includegraphics[width=\linewidth]{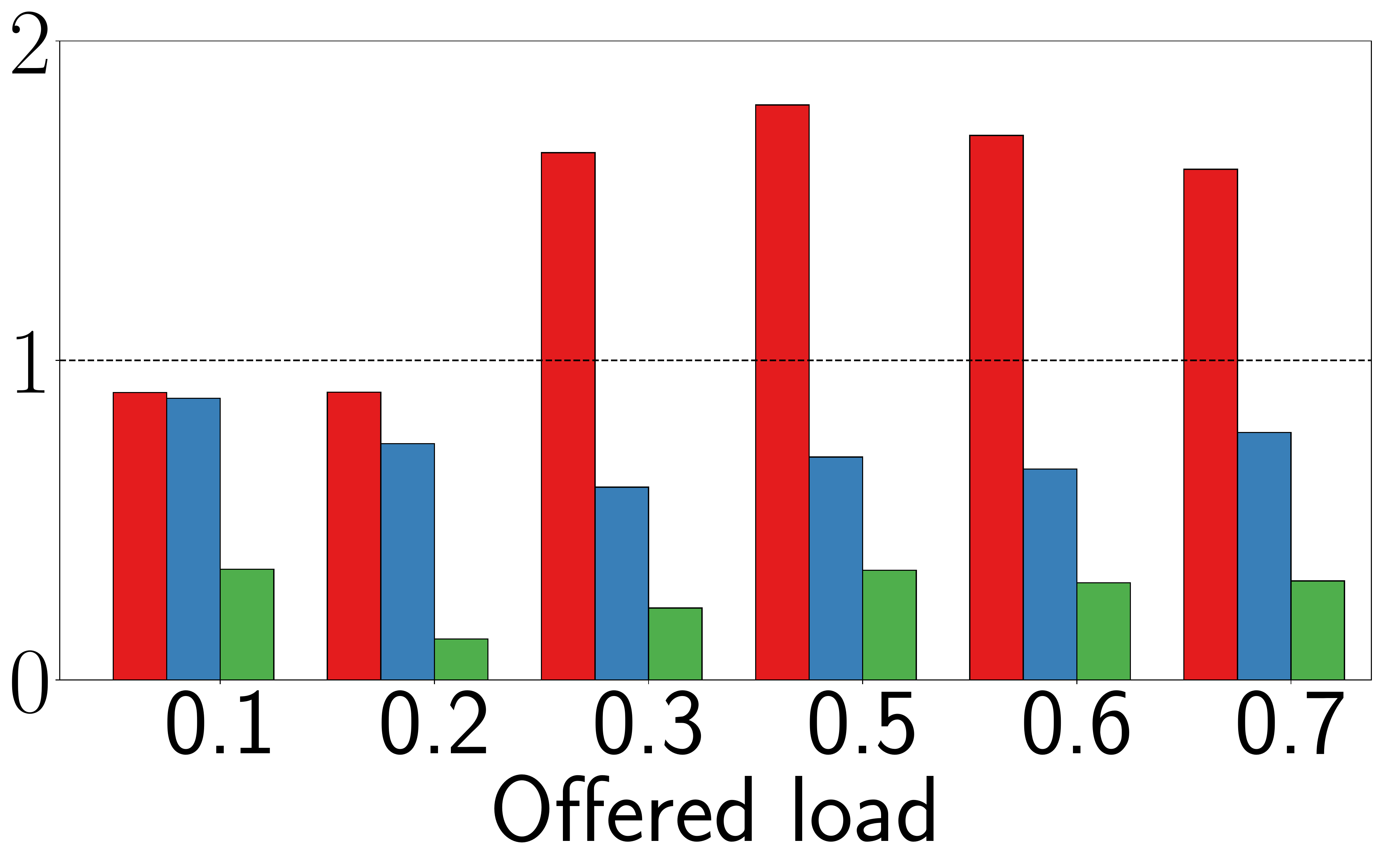}
  \caption{Transpose.}
  \label{fig:offered_load_transpose_ugal} 
  \end{subfigure}  
\vspace*{-1em} \\
\centering
\includegraphics[width=0.4\textwidth]{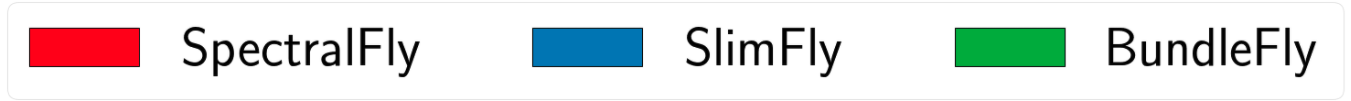}
\vspace{-1.0em}
\caption{
Performance comparison across topologies, traffic patterns, and offered load conditions under UGAL-L routing. \vspace{-1.5em}
}
  \label{fig:offered_load_ugal}
\end{figure*}

\section{Simulation Results} \label{sec:sim}
In this section, we report our simulation results on evaluating SpectralFly,  
SlimFly, BundleFly and DragonFly topologies with different workloads exhibiting interesting communication patterns that are prevalent in many HPC applications.  

\subsection{Simulation Software}
We conduct our experiments in the Structural Simulation Toolkit (SST) Macroscale Element Library (SST/macro) simulator~\cite{sstrepo}.
Our simulation approach performs online simulation, which involves skeletonization of an application during the compilation step so that part of the application involving communication (such as communication API calls, MPI\_alltoall etc.) can be intercepted by the simulator during runtime.  The simulator replaces these calls with various built-in network component model implementations. The application can then run inside the simulator without any significant change. The user can provide necessary hardware parameter values (for routers, NICS, topologies, routing schemes etc.) to the simulator for running the application with different hardware configurations. 
We have used the \textit{Simulator Network for Adaptive Priority Packet Routing (SNAPPR)} network model in SST/macro to evaluate different topologies. SNAPPR  implements coarse-grained cycle-based simulation to simulate priority queue-based QoS. In addition, it can also restrict injection rate of messages for congestion control. For a detailed discussion about available network models in SST/macro, we refer to the SST/macro user manual~\cite{sstrepo}.

\subsection{Configuration and Simulation Setup}\label{subsec:config}

We evaluate the performance of 
different micro-benchmarks  by considering SpectralFly, DragonFly, SlimFly and BundleFly topologies. We conducted our experiments with \textapprox{}$8.7k$ network endpoints and with 32-port routers. To generate the SpectralFly topology with \textapprox{}$8.7k$ network endpoints, we set $(p, q)=(23, 13)$ to generate a graph with 1092 routers, and a  concentration of $8$ endpoints per router. 
For the DragonFly topology configuration, the number of groups used is 69 ($g$), with 16 routers per group ($a$), each router connected to $8$ endpoints ($p$), and 8 global links ($h$) per router. This conforms to the recommended balance to support full global bandwidth for Dragonfly with radix-$k$ switches ($p=k/4, h=k/4, a=k/2$). The global links in the DragonFly topology are arranged in a circulant manner~\cite{hastings2015comparing,kaplan2017unveiling}, since this arrangement provides better bisection bandwidth than the absolute arrangement.
For the SlimFly topology, $q$ is set to 27, with each router connected to 8 endpoints. Finally, for the BundleFly topology, the graph is constructed with $p=s=9$, and each router has a concentration of 6 endpoints. 

In the case of under-subscription (for example, when running microbenchmarks with $2^{13} = 8192$ ranks out of \textapprox{}$8.7k$ available ranks), \change{the physical nodes allocated to the job} are chosen randomly. 
\change{Each MPI rank is then sequentially allocated to nodes based on the standard ordering for the topology.  For the SpectralFly topology we use the essentially unstructured ordering resulting from the Elzinga construction~\cite{elzinga2010producing}.}
We report our experimental results with various routing strategies.  Valiant routing demonstrates similar performance trend. The router buffer size has been set to 64KB (Other buffer sizes have also been tested but the results are not reported here due to space constraint). For simulation, the number of virtual channels has been set to the diameter of the graph plus one.

\subsection{Experimental results}

\begin{figure}[tp]

    \centering
  \includegraphics[width=0.80\linewidth]{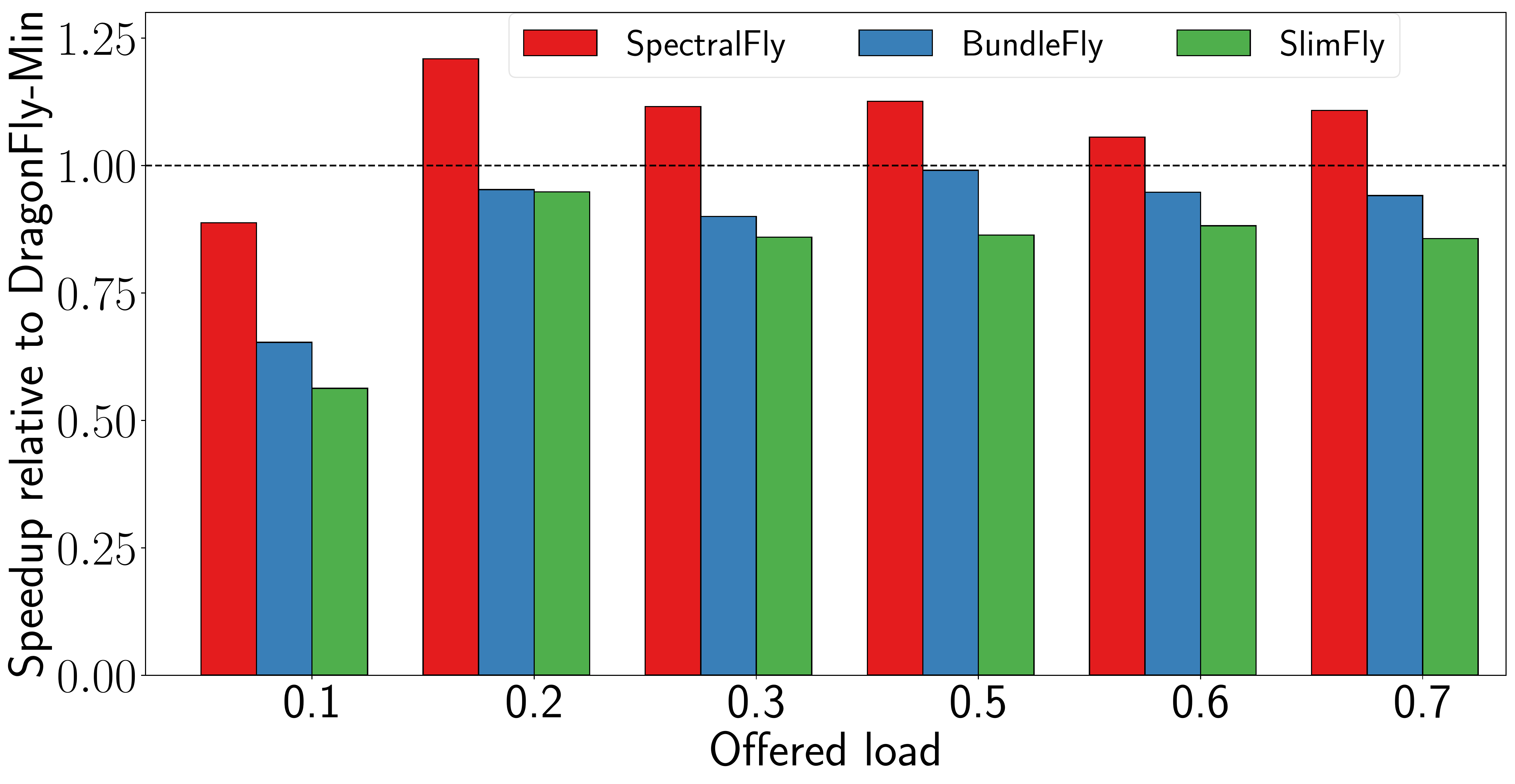}
  \label{fig:offered_load_random}
\vspace{-0.8em}
\caption{
Performance across topologies, 
and offered load conditions with random micro-benchmark, under minimal routing. \vspace{-3em}}
  \label{fig:offered_load_minimal}
\end{figure}

\subsubsection{Micro-benchmarks to assess performance under congestion}
We consider standard traffic pattern micro-benchmarks to evaluate the performance of different topologies under various network capacities (offered load). These include random, bit shuffle,  transpose, and bit reverse traffic patterns. In each case, a source node communicates with a destination node that is determined by a specific permutation of the bit representation of the source.  Random traffic patterns can be found in many irregular and graph applications. The shuffle traffic pattern (obtained by rotating left 1 bit of the source) can be found in Fast Fourier Transform (FFT) and sorting applications. Matrix transpose is a basic linear-algebraic operation. 

We consider a total of 8192 endpoints for these experiments. For each traffic pattern ran on a topology, we collect the maximum time taken across all the messages under a particular offered load. 
The results are reported in~\Cref{fig:offered_load_ugal}. Here, on the x-axis we plot the offered load i.e. how much of the network is saturated when running the micro-benchmarks. To simulate network congestion, we inject messages with varying delays by simulating a Poisson process. We report the speedup relative to the execution with 
DragonFly 
Each topology was run with the  UGAL-L routing. As can be observed from the figure, for all the micro-benchmarks 
SpectralFly performs the best.
The better performance of SpectralFly  can be attributed to the superior bisection bandwidth and available path diversity of the SpectralFly topology.
At or beyond 70\% of the network capacity, the network becomes saturated.
Between BundleFly and SlimFly, BundleFly exhibits better performance (except with bit shuffle traffic). These experiments demonstrate that, because of stronger discrepancy and spectral properties, SpectralFly 
is robust to accommodate diverse traffic patterns under varying degrees of network congestion.

\begin{figure}[tp]
  \centering
  \includegraphics[width=0.7\linewidth]{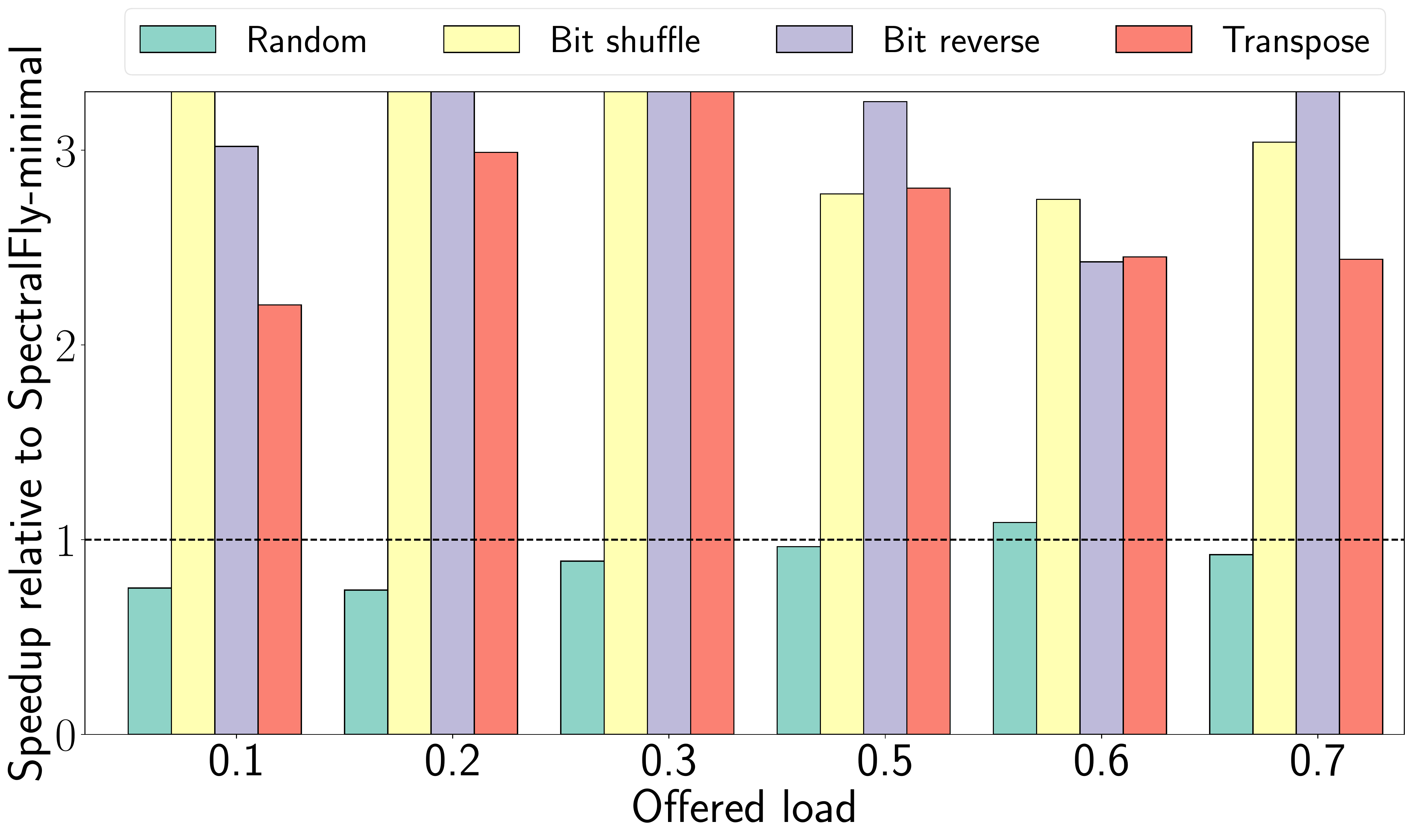}
\vspace{-0.8em}
 \caption{
 \change{Evaluation of Valiant routing for the SpectralFly topology with micro-benchmarks.\vspace{-2em}}}
  \label{fig:offered_load_random_lps_routing}
\end{figure}

\subsubsection{Evaluation of different 
routing schemes}
Besides evaluating different topologies with the UGAL routing, we also consider minimal and Valiant routing for evaluation. \Cref{fig:offered_load_minimal} presents the performance of the random 
benchmark with minimal routing. With random micro-benchmark, SpectralFly demonstrates better performance, compared to other topologies. Bit shuffle and transpose exhibit similar patterns. Next, in order to evaluate the difference between minimal and Valiant routing for SpectralFly, we ran the four micro-benchmarks under varying network loads  and report the results in \Cref{fig:offered_load_random_lps_routing}.  The execution time is normalized w.r.t. the execution time of minimal (shortest-path) routing scheme on SpectralFly.

We see significant improvement for all offered loads with the bit shuffle, bit reverse, and transpose traffic patterns, while the random pattern has a significant decrease in performance (except at 60\% offered load).  This suggests the increase in path diversity by applying Valiant routing to a structured communication pattern better exploits the discrepancy property of the LPS graphs.  Moreover, there is already significant path diversity in minimal routing for the random micro-benchmark, and so the addition of Valiant routing provides a minimal increase in path diversity while doubling the expected length of the routing paths.  This suggests SpectralFly performs best when traffic is unstructured, either due to the logical communication pattern or from the choice of routing algorithm.

{
\subsection{Evaluation of Topologies under Real-World Traffic Patterns}

\subsubsection{Patterns Considered}
For evaluating different topologies with real-world traffic patterns (under both minimal and UGAL routings), 
we consider communication motifs from the Ember Communication Pattern Library~\cite{emberrepo}: 
\begin{enumerate}[label=(\roman*),wide, labelwidth=!, labelindent=0pt]

\item \textit{Nearest neighbor communication pattern -- Halo3D-26:} 
Nearest neighbor communication pattern, found in stencil workloads, is captured by the Halo3D-26 motif, where each MPI rank communicates with 6 of it's nearest neighbors as well as 20 of it's diagonal neighbors, 
a total of 26 neighbors. 

\item \textit{Wavefront communication pattern -- Sweep3D:} Wavefront communication pattern is prevalent in particle transport physics simulation, parallel iterative solvers and  triangular solvers \cite{hoisie1999performance} that generally stresses network latency and has substantial dependency levels. One representative motif for the wavefront communication pattern is the ASCI ~Sweep3D~\cite{hoisie1999performance}. 
Here, a 3D data domain is decomposed over a 2D array of MPI processes and repeated sweep along the diagonal is performed. 

\item \textit{Subcommunicator all-to-all communication pattern-- FFT:} In this communication pattern, found in Multi-dimensional FFT, a 3D domain is decomposed along the $X$ and $Y$ dimensions and subcommunicators are formed along the 1D line in both of the $X$ and $Y$ dimensions. One MPI process is assigned to each of the 3D grid points and communicates with all the subcommunicators along 
the $X$ and $Y$ dimensions.
\end{enumerate}
\begin{figure}[tp]
  \centering
  \includegraphics[width=0.7\linewidth]{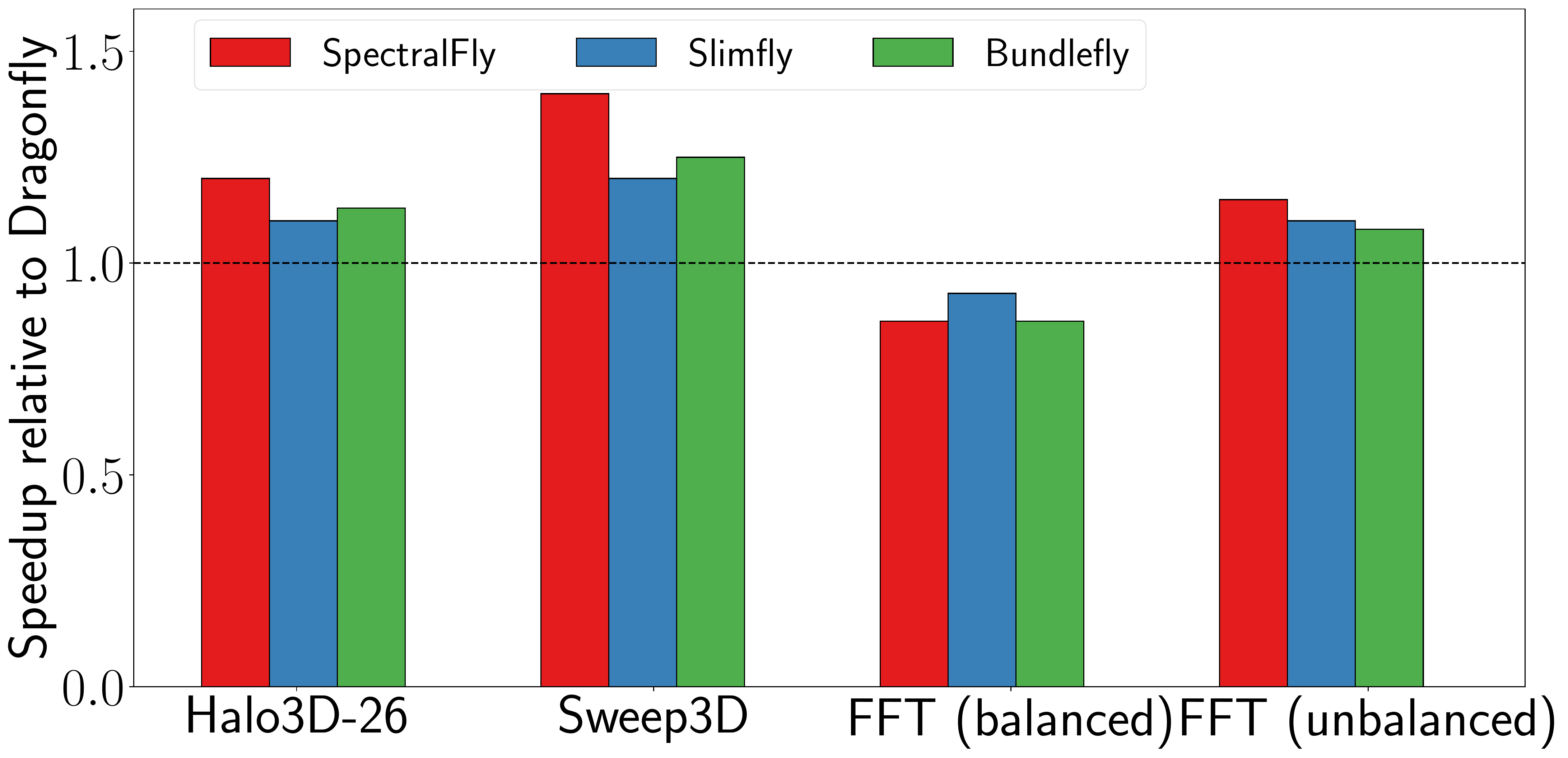}
 \vspace{-1.2em}
 \caption{\small{Performance under Ember real-world traffic patterns with minimal routing, reported as speedup w.r.t. the DragonFly topology. \vspace{-1.5em} } }
  \label{fig:ember_speedup}
\end{figure}
\begin{figure}[tp]
\centering
  \begin{subfigure}[b]{\columnwidth}
    \centering
  \includegraphics[width=0.7\columnwidth]{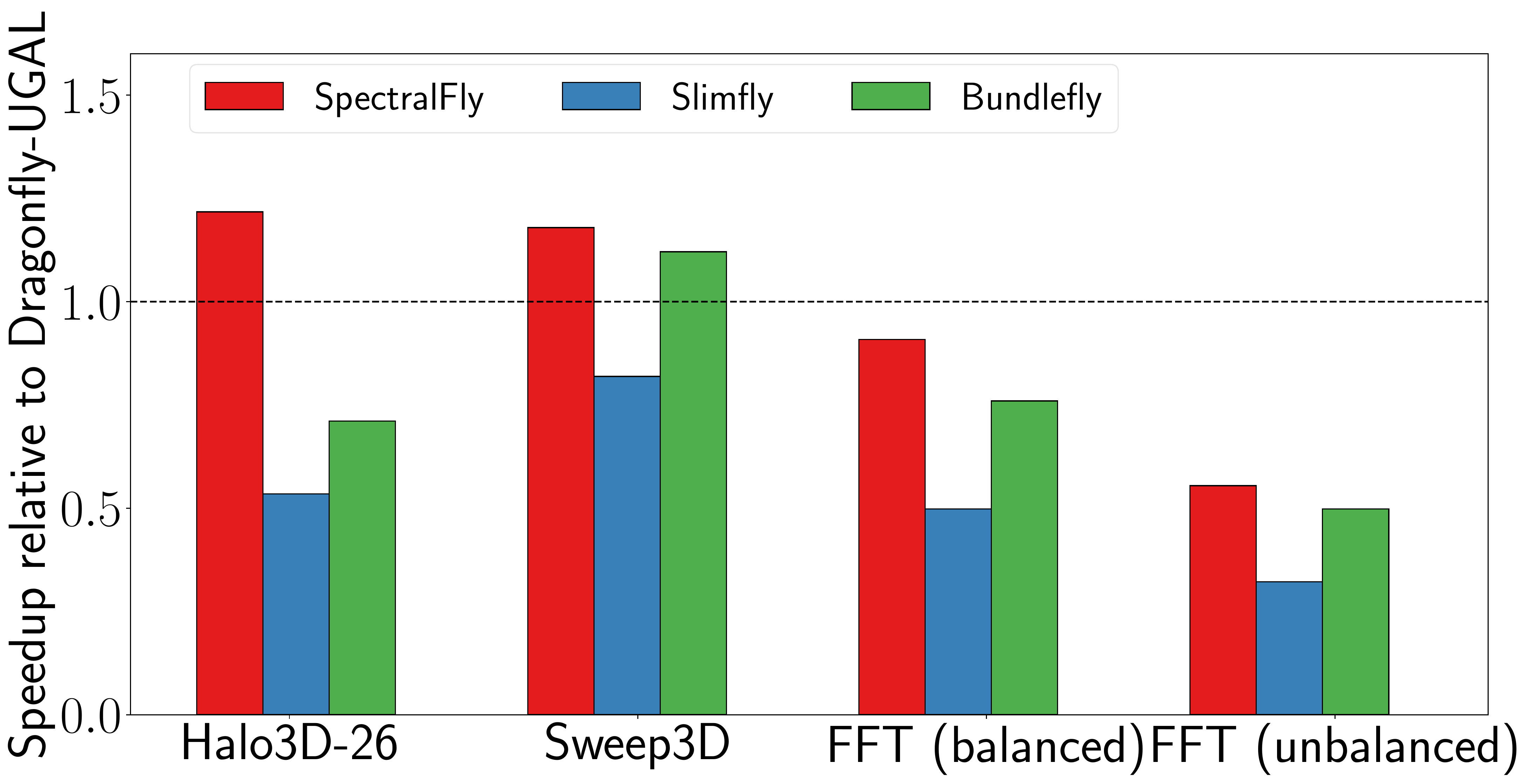}
  \end{subfigure} \\ 
\vspace{-1.2em}
\caption{\change{
\small{Performance of different topologies with UGAL routing w.r.t. DragonFly UGAL routing for real-world traffic patterns. \vspace{-1.5em}}}}
  \label{fig:ember_ugal_speedup}
\end{figure}

\subsubsection{Performance Results}
The performance of each of the Ember motifs on different topologies are reported in~\Cref{fig:ember_speedup} (with minimal routing) and \cref{fig:ember_ugal_speedup} (with UGAL routing).
As can be observed from~\Cref{fig:ember_speedup}, for both the Halo3D-26 and the Sweep3D traffic patterns, the SpectralFly configuration outperforms the other topologies with a speedup of \textapprox{}$1.2\times$ and \textapprox{}$1.4\times$ respectively, over the DragonFly topology (with minimal routing).  This indicates that, for SpectralFly, with communication patterns with relatively low per-node communication, the robust discrepancy property and the reduction in the average hop-count is sufficient to ameliorate any penalty accruing as a result of longer maximum hop-count.  In contrast to this, we see that for the balanced FFT motif, DragonFly slightly outperforms the other topologies. As the communication pattern for FFT involves all-to-all communication along a 2D-plane within a 3D-arrangement of ranks, we suspect that relative improvement is a result of the partial alignment of these 2D all-to-all communication with the group structure.  Specifically, if multiple nodes from the same all-to-all communication land in the same group, there is an out-sized decrease in the communication pressure on the global links. In particular, we note that the stronger group structure of DragonFly (even as compared to BundleFly and SlimFly) leads to the best performance on the balanced FFT motif.  We also note that, because of the lack of large all-to-all clusters in Halo3D-26 and Sweep3D, there is as much marginal benefit to alignment with the group structure.  Finally, for the unbalanced FFT traffic pattern, the SpectralFly configuration outperforms all other topologies.  
While the other topologies with strong group structure 
will again benefit from multiple elements from the 2D all-to-all aligning with the group, the increased sizes of the all-to-all groups will necessitate significantly more between-group traffic on global links, degrading overall performance.  In contrast, 
SpectralFly 
handles the increased all-to-all communication pressure better. 

We also evaluate the Ember benchmarks on different topologies with UGAL routing. \Cref{fig:ember_ugal_speedup} shows SpectralFly outperforms other topologies for halo3D-26 and Sweep3D motifs. However, for both FFT motifs,  
DragonFly with UGAL routing performs better.
For the FFT motif, SpectralFly performs better than SlimFly and BundleFly (achieving 90\% of the execution efficiency w.r.t DragonFly for the balanced FFT motif). 
This suggests discrepancy properties of LPS graphs ensure performance of SpectralFly either better-than or competitive-with topologies which excel under the UGAL routing scheme. 
}

\begin{table*}[t]
{
\footnotesize
\centering
\begin{tabular}{l|c|c|cc|cc|c|c |c| c|c}
\multirow{2}{*}{\bf Topology} & \multirow{2}{*}{\bf Routers} & {\bf Router} & \multicolumn{2}{|c|}{\bf Average Wire} & \multicolumn{2}{|c|}{\bf Max.\ Wire} & {\bf Electrical} & {\bf Optical} & {\bf Bisection} & {\bf Total} & {\bf Power/Bandwidth} \\
& & {\bf Radix} & \multicolumn{2}{|c|}{\bf Length (m)} &\multicolumn{2}{|c|}{\bf Length (m)} & {\bf Links} & {\bf Links} &{\bf Bandwidth} & {\bf Power (W)} & {\bf (mW per Gb/s)} \\
     \toprule
     LPS$(11,7)$ & \phantom{0}168 & 12 & \phantom{0}8.02 & \multirow{2}{*}{(10.29)} & 19.8 &\multirow{2}{*}{(21.21)}& 249 & \phantom{00}758 & 304 &\phantom{00}928 & 30.5\\
    SF$(9)$ &  \phantom{0}162 &    13 & \phantom{0}8.68 && 21.6 && 151 & \phantom{00}902 & 369 & \phantom{0}1028 &27.9\\
    \hline
    LPS$(19,7)$ & \phantom{0}336 & 20 & 10.43 &\multirow{2}{*}{(13.94)}& 28.6 &\multirow{2}{*}{(31.05)}& 432 & \phantom{0}2928  &1080& \phantom{0}3276& 30.3\\
    SF$(13)$ &  \phantom{0}338 & 19&  10.89 && 27.8 && 315 & \phantom{0}2896 & 1105 & \phantom{0}3155 & 28.6\\
    \hline
    LPS$(23,11)$ & \phantom{0}660 & 24 & 14.35 &\multirow{2}{*}{(17.27)}& 39.8 &\multirow{2}{*}{(41.07)}& 531 & \phantom{0}7389 & 2928 & \phantom{0}7845& 26.8\\
    SF$(17)$ & \phantom{0}578  & 25 & 13.05 && 36.2 && 558 & \phantom{0}6667 & 2465 & \phantom{0}7138 & 29.0\\
    \hline
    LPS$(29,13)$ &  1092 & 30 & 17.32 &\multirow{2}{*}{(21.09)}& 50.8 &\multirow{2}{*}{(52.10)}& 831 & 15549  & 6150 & 16292& 26.5\\
    SF$(23)$ & 1058 &  35 & 16.00 && 47.4 && 1257 & 17258  & 6095 & 18336& 30.1\\
    \bottomrule
\end{tabular}
\caption{\small Wire length and energy efficiency statistics for the heuristic embedding of comparable SpectralFly and SlimFly topologies.  For mean and maximum wire length, we include in parenthesis the mean 20 instantiations of the SkyWalk topology in the same machine room. \vspace*{-2.0em}}\label{tab:wirelengths}}
\end{table*}

\begin{figure}
    \centering
    \includegraphics[trim = 15 15 0 15,clip,width = \linewidth]{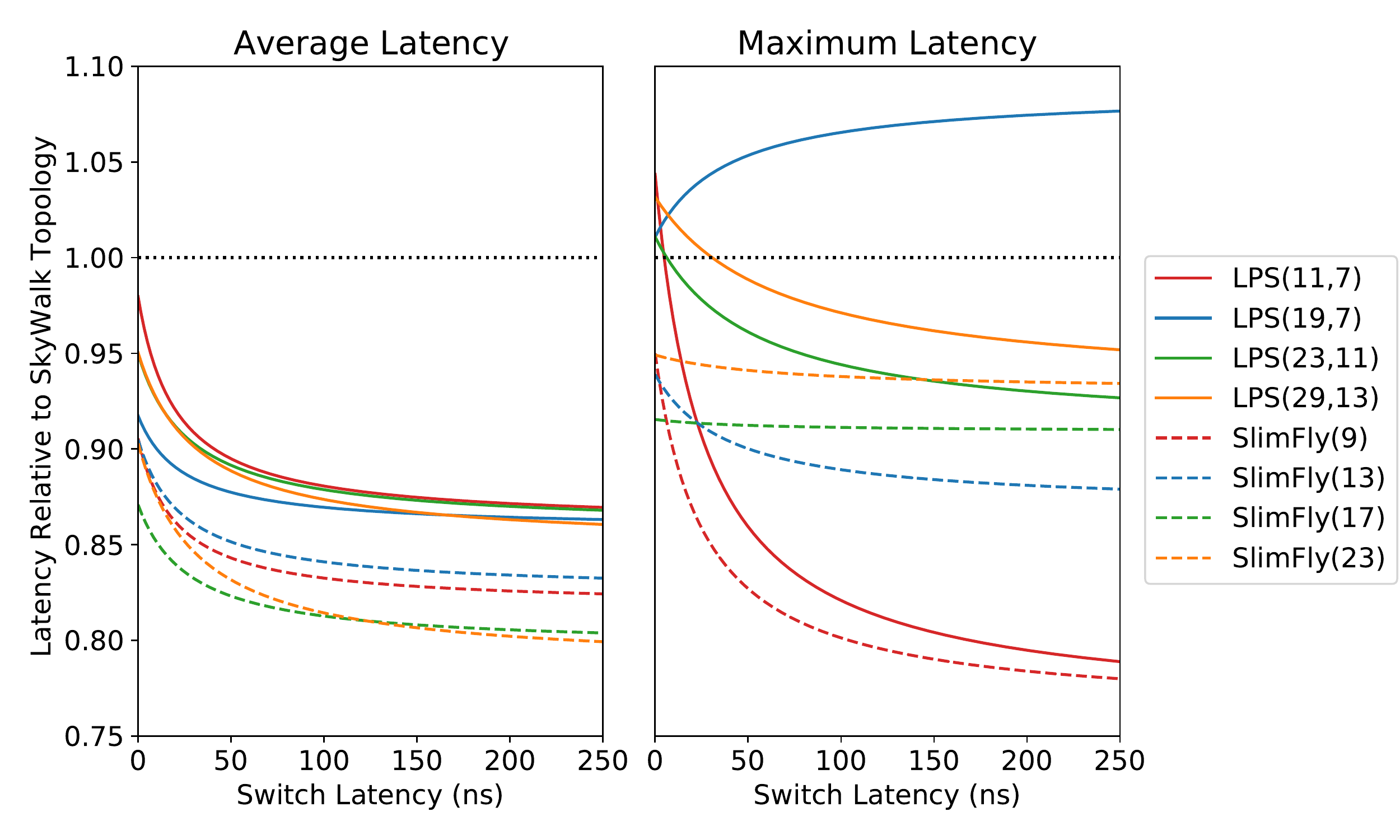}
    \caption{\small Ratio of maximum and average latency between SpectralFly/SlimFly and SkyWalk as a function of the switch latency. \vspace{-3em}}
    \label{fig:latency}
\end{figure}
\vspace{-.25em}
\section{Beyond Structure}
As noted in Section \ref{sec:structProp}, network parameters for each topology (Table \ref{tab:structProp}), were chosen to facilitate a comparison of interconnects based on their fundamental \emph{structural} properties.  In Sections \ref{sec:structProp} and \ref{sec:sim}, the structure of SpectralFly is superior to, or comparable with, that of DragonFly, SlimFly, and BundleFly across a variety of metrics.  However, in practice, the trade-off between topology cost and performance is an important factor in the overall design. 
 Since the competing topologies have a similar number of routers and connections per router, the total amount of wiring needed to build the topologies will be a primary driver of any cost differences. {In addition to the direct costs of wire length there is an additional, indirect cost as longer wires often necessitate higher energy optical connections. As an added benefit, the analysis of wire lengths allows us to evaluate the end-to-end and typical latency of SpectralFly as compared to physical latency minimizing topologies, such as SkyWalk~\cite{SkyWalk}.}

First, we compare the average {and maximum} wire length necessary to implement a SpectralFly topology to the similarly-sized SlimFly topology. SlimFly was chosen as point of comparison because the bisection bandwidth (see Figure \ref{fig:sizes_bws}) is most structurally comparable to the SpectralFly topology. To ensure an equitable comparison, we assume each topology is implemented equal concentration with rectilinear physical wiring. {Following the methodology in \cite{SkyWalk}, we assume an $x \times y$ grid of cabinets where intra-cabinent wires are all $2$ meters while the inter-cabinet wires have length of $4 + 2 \abs{x_i-x_j} + 0.6 \abs{y_i - y_j}$, which includes a 2 meter overhead at each end of the link. Assuming a roughly square room, we fix $y = \lceil\sqrt{\nicefrac{2c}{0.6}}\rceil$ and $x = \lceil\nicefrac{c}{y}\rceil$ where $c$ is the minimum number of cabinets need for the topology if, similar to the Summit supercomputer,  each cabinet contains two routers.}

This allows us to restrict our attention to the wiring between the routers. Thus the question of minimal average wire length is an instance of the Quadratic Assignment Problem (QAP), which is $\mathcal{NP}$-complete.  To find a heuristically minimal layout, we apply an expectation minimization approach combined with a greedy refinement process which outperforms the the standard Fast Approximate QAP algorithm on these instances~\cite{Vogelstein:FAQ}. {In order to take advantage of short lengths of intra-cabinent links, for each topology we fix as a maximum matching of the underlying topology and enforce that the matching edges are within a cabinet.  Table \ref{tab:wirelengths} provides a summary of the results of this layout approach. As we can see the maximum and average wire lengths of SpectralFly and SlimFly topologies are within \textapprox 10\% of each other across all sizes, with SpectralFly performing better on smaller topologies.  To provide additional context, we compare the layout with the SkyWalk topology which was designed to minimize end-to-end latency in the case of ultra-low latency routers/switches.  For each machine room, we report (in parenthesis) the average behavior over 20 instantiations of the SkyWalk topology in the same machine room.  As we can see the SkyWalk topology typically requires between \textapprox{}20-30\% longer lines overall with a maximum wirelength \textapprox{}3\% longer.  This indicates that despite the underlying expansion of the SpectralFly and SlimFly necessitating longer wires, with care the overall wire lengths can be made comparable to other modern topologies.}

{
To translate the wire lengths to an estimate of the power usage, we update the methodology of \cite{GooglePower} to modern hardware (i.e., the Mellanox SB7800
InfiniBand EDR 100Gb/s Switch) and assume each port connected to an electrical link uses \textapprox{}3.76 W while ports with optical links use 25\% more power at \textapprox{}4.72 W.  Using METIS to approximate bisection bandwidth, we quantify the trade-off between overall power expenditure versus communication performance (see Table \ref{tab:wirelengths}). As is the case with other metrics, the difference in energy cost per unit of bandwidth is \textapprox{}5-10\%, with the notable exception of the $(29,13)$-SpectralFy being 15\% more efficient than the similarly sized SlimFly.  This efficiency gain is a consequence of SpectralFly's better expansion properties yielding slightly better bisection bandwidth while requiring \textapprox{}15\% fewer links.}

{
The wire lengths allows the evaluation of end-to-end latency and clock cycle times implicit in SpectralFly and SlimFly.  Following \cite{SkyWalk}, we assume a cable delay of $5\ \nicefrac{ns}{m}$ uniform switch latencies.  Figure \ref{fig:latency} provides a comparison of both SlimFly and SpectralFly with the latency minimizing SkyWalk topology. Except for $\mbox{LPS}(19,7)$, we have that both topologies typically have lower end-to-end latency (and hence clock-cycle time) than the SkyWalk topology, as well as singificantly lower average latency.   While the average and end-to-end latency of SpectralFly is slightly larger (\textapprox{}5-10\%) than SlimFly, necssitating a longt clock-cycle time, the overall performance  benefits illustrated in Section \ref{sec:sim} are sufficient to make up for this difference.  Further, we believe that applying a more sophisticated multi-objective minimization approach to the layout problem will further close the gap in latencies between these two topologies.
}

\section{Conclusion}

The design of interconnection networks is increasingly informed by graph theoretic considerations. While researchers have established a long list of desirable criteria, such as low-diameter and average distance, high fault tolerance, and high bisection bandwidth, developing topologies exhibiting all these properties requires sophisticated methods. To this end, we've proposed SpectralFly, a class of topologies with optimal spectral gap based on the LPS graph algebraic construction.  

Exploring first the design space of LPS graphs, we showed this construction permits a large range of topology sizes and radix values, including arbitrarily large topologies per fixed radix. We then highlighted, both via experiments and analytically, structural properties for which SpectralFly excelled in comparison to competing topologies. In particular, for bottleneck measures such as normalized bisection bandwidth, SpectralFly outperformed other topologies, which have decaying or tightly bounded bandwidth. The concession for these properties is slightly larger diameter; however, we showed the average distance between nodes in an LPS topology is typically smaller than DragonFly and BundleFly, and only marginally larger than SlimFly. Furthermore, experiments suggest these LPS graph properties remain relatively robust under edge failures. Lastly, in order to experimentally validate the potential of SpectralFly suggested by its structural properties, we conducted 
simulations using the SST/macro simulator. SpectralFly outperformed other network topologies under a diverse range of communication patterns found in traditional HPC workloads.  Further, we demonstrated that the cost of implementing a SpectralFly topology is on-par with, if not better than, the SlimFly topology (which is the only considered topology which has comparable bandwidth). \\

\noindent {\bf Acknowledgement.} We would like to thank Jeremiah Wilke for very helpful technical exchanges regarding SST/macro. This work was supported by the High Performance Data Analytics program at PNNL. Information Release PNNL-SA-160551.

\bibliographystyle{IEEEtran}
\bibliography{IEEEabrv,references}
\end{document}